\documentclass[12pt]{article}

\usepackage[a4paper, margin=1in]{geometry}
\usepackage{setspace}  
\usepackage{caption}
\usepackage{lmodern}     
\usepackage{authblk} 
\usepackage[colorlinks=true, linkcolor=blue, citecolor=cyan, urlcolor=blue]{hyperref} 
\usepackage{graphicx}
\usepackage{float}
\usepackage{amssymb,amsmath,mathtools,amsfonts}
\evensidemargin\oddsidemargin \hoffset-22.4mm \voffset-28.4mm
\usepackage[numbers,sort&compress]{natbib}
\usepackage{titlesec}

\begin{document}

\title{How localized nonlinear losses condition the acoustical design of a self-sustained oscillator: the clarinet and its register hole}
\author[, a, b]{Nathan Szwarcberg\thanks{Corresponding author. \\
E-mail address: nathan.szwarcberg@gmail.com (N.\ Szwarcberg)}}
\author[a]{Tom Colinot}
\author[b]{Christophe Vergez}
\author[a]{Michaël Jousserand}

\affil[a]{Buffet Crampon, 5 Rue Maurice Berteaux, 78711 Mantes-la-Ville, France}
\affil[b]{Aix Marseille Univ, CNRS, Centrale Med, LMA, Marseille, France}

\maketitle
\begin{abstract}
The register tube marks the invention of the clarinet in the early eighteenth century, tripling the range of its ancestor, the \textit{chalumeau}, and giving it the widest range among wind instruments.
Opening this narrow tube causes the fundamental frequency of the played note to increase by a factor of three, from the first to the second register of the resonator.
The geometry and location of the register hole condition not only this mode selection mechanism, but also the global tuning of the second register.
However, existing self-sustained nonlinear models of reed instruments fail to predict whether a register transition can occur, limiting optimization of the register hole geometry.
Here, we introduce a sparse self-oscillating clarinet model that includes localized nonlinear acoustic losses in the register hole.
This nonlinear mechanism is shown to be necessary to reproduce register transitions observed experimentally.
Using systematic exploration of the control and design parameter spaces, we identify combinations of register hole diameter, position, and chimney length that ensure reliable register transitions.
We show that the competing demands of \textit{playability} and tuning are only satisfied by a long and narrow tube.
Our findings provide a predictive tool for instrument making, assisting manufacturers in refining clarinets as well as other reed instruments, including oboes, bassoons, and saxophones.\\

\end{abstract}

\section{Introduction}

The invention of the clarinet is attributed to the German instrument maker J.~C.~Denner, in the beginning of the eighteenth century \cite{lawson1995cambridge,hoeprich2008clarinet,rice2020baroque}. 
It derives from the \textit{chalumeau}, a single-reed instrument with a cylindrical bore and tone holes. 
The chalumeau had a limited playing range of about a musical tenth, within a single register.
The standard range of modern clarinet is almost three times as wide.

\begin{figure}[H]
	\centering
	\includegraphics[width=\textwidth]{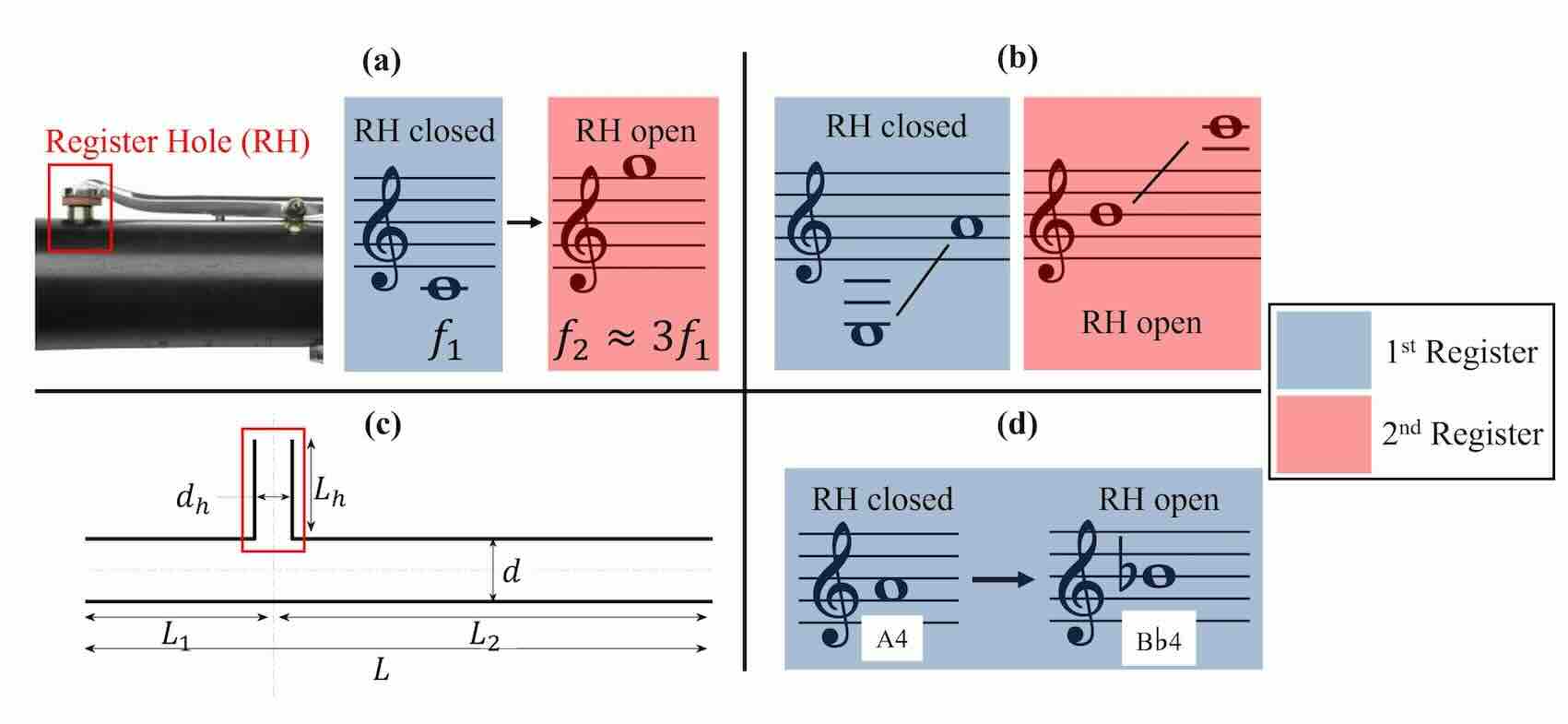}
	\caption[Overview of the operation of the register hole of the clarinet.]{Overview of the operation of the register hole (RH) of the clarinet.
	(a): \textit{Overblowing} from the first register (in blue, at frequency $f_1$) to the second register (in red, at frequency $f_2$).
	(b): Playing range of first-register and second-register notes of the clarinet.
	(c): Schematic view of a clarinet resonator featuring a RH.
	(d): The \textit{throat} B$\flat$4, bridging the gap between the top of the first register (A4) and the bottom of the second register (B4).}
	\label{chap5:fig:0} 
\end{figure}

Denner’s decisive innovation was the introduction of a \textit{register hole} (RH), shown in Figure~\ref{chap5:fig:0}(a).  
Opening the RH allows the instrument to \textit{overblow}, moving from the first register (the chalumeau range, at frequency $f_1$) to the second register (at $f_2 \approx 3 f_1$).  
This overblowing mechanism works for all fingerings from the low E3 to the F4. 
Because this jump is large, the top of the first register (A4\footnote{All notes are written in B$\flat$ in this article. The frequency of B4 is therefore $440$~Hz.}) and the bottom of the second register (natural B4) are separated by one tone [Fig.~\ref{chap5:fig:0}(b)].  
To bridge this gap, the RH also acts as a tone hole and raises the pitch of A4 by a semitone, producing the \textit{throat}~B$\flat$4 [Fig.~\ref{chap5:fig:0}(d)].

Unlike an ordinary tone hole, the RH is not a simple drilling, but a narrow metal tube inserted into the wood.
Various hypotheses have been proposed for this distinctive design. 
Mechanically, the metal tube may prevent moisture intrusion \cite{greenham2003clarinet,rice2020baroque}. 
Acoustically, linear models based on resonance frequencies show that a longer chimney reduces the \textit{inharmonicity} $h$ between the first and second registers,
\begin{equation}
h = \frac{f_2^{(o)} - 3 f_1^{(c)}}{3 f_1^{(c)}},
\end{equation}
where $f_1^{(c)}$ is the first resonance frequency with the RH closed, and $f_2^{(o)}$ is the second resonance frequency with the RH open.  
Using the classical \textit{length corrections} approach, the inharmonicity introduced by an open RH in an ideal lossless cylindrical tube is
\cite{benade1960mathematical,nederveen1969acoustical,hoekje1988abriefsummary,debut_analysis_2005}:
\begin{equation}\label{chap5:eq:h}
h \approx \frac{c_0}{9 \pi^2} \frac{1}{L_h} \frac{\phi_2^2(L_1)}{f_1^{(c)}} \left( \frac{d_h}{d} \right)^2,
\end{equation}
where $c_0=343$~m/s is the speed of sound in the air, $L_h$ the RH chimney length, $d_h$ the RH diameter, $d$ the main bore diameter, and $L_1$ the distance from the the mouthpiece tip to the RH.
The second modeshape $\phi_2$ of a closed-open tube is
$$ \phi_2(x) = \cos \left( \frac{3 \pi}{2} \frac{x}{L} \right), $$
where $L$ is the total length of the instrument [Fig.~ \ref{chap5:fig:0}(c)].

This linear approach predicts how inharmonicity varies with geometry and remains helpful for instrument makers \cite{debut_analysis_2005}.  
Yet, it cannot determine whether a given RH actually enables a successful overblowing. 
For example, a B$\flat$ clarinet usually has $d_h \approx 3.0$~mm \cite{debut_analysis_2005}.  
However, playing tests with musicians show that for $d_h = 1.0$~mm, a clarinet prototype remains locked in the first register even when the RH is open \cite{szwarcberg2024second}.  
Hence, reducing $d_h$ lowers inharmonicity, but can also prevent the first register from losing its stability.

This limitation is also present in classical self-oscillating models.  
These nonlinear models often predict a stable first register even for fingerings of the second register \cite{scavone1997acoustic,szwarcberg2024second,colinot2025cartography}.

At the same time, experimental evidences show that localized nonlinear dissipation can occur at geometric discontinuities in wind instruments \cite{keefe1982experiments,blevins1984review,dalmont_experimental_2002,atig2004termination,atig2004saturation,buick2011investigation, lasickas2025investigating}. 
Large particle displacement produces \textit{vortex shedding} at these discontinuities \cite{fabre1996vortex,martinez2023generation}.  
This effect can be modeled either as a \textit{nonlinear acoustic impedance} \cite{ingaard1950acoustic,dalmont_experimental_2002,atig2004termination,temiz2016non,diab2022nonlinear} or via a quasi-steady Bernoulli-type boundary condition \cite{disselhorst1980flow,cummings1983high,ducasse1990modelisation,atig2004saturation,hirschberg2004introduction,guillemain2006digital,temiz2016non,szwarcberg2025oscillation}.
In a previous communication \cite{szwarcberg2024second}, localized nonlinear losses in the RH helped restore register transitions when the geometry and position of the RH matched the experimental prototype.  
However, if the geometry of the prototype is not accurately reproduced numerically, the model may fail to reproduce the observed transitions.

To date, no numerical approach has been able to predict RH dimensions that guarantee reliable transitions to the second register.  
In this study, we take one step further by integrating practical constraints from instrument making.  
We explore the set of RH dimensions that allow a \textit{playable} instrument from the bottom of the first register (E3) to the top of the second register (C6).
We define playability through three criteria.  
The first one is the ability to overblow from E3 to F4 [Fig.~\ref{chap5:fig:0}(a)].  
The second one is the minimization of inharmonicity between the first and second registers.  
The third one is the ability to produce the throat B$\flat$4 when opening the RH from A4 [Fig.~\ref{chap5:fig:0}(d)].

We first show that localized nonlinear losses in the RH are essential to trigger a transition to the second register.  
On the experimental side, we use a clarinet prototype with interchangeable RHs (Section~\ref{chap5:sec:exp}).  
Playing tests with clarinetists evaluate the ability of each RH to overblow.  
We then introduce a sparse waveguide model with localized nonlinear losses (Section~\ref{chap5:sec:model}).  
Results in Section~\ref{chap5:sec:results:1} show that the model captures register transitions reliably.  
We then identify a set of RH dimensions that meets the playability criteria and reveal strong design tensions (Section~\ref{chap5:sec:results:2}).  
Under these constraints, we show that only a long-chimney RH can correct the severe tuning issues of the lowest notes (Section~\ref{chap5:sec:results:3}).  
We conclude by proposing a revised design that improves tuning across the second register while preserving playability.

\section{Material and Methods}

This section introduces the experimental setup and numerical model considered in this article.
Clarinet-like prototypes are first presented in Section~\ref{chap5:sec:exp:res}. 
They are characterized experimentally through a playing-test process, defined in Section~\ref{chap5:sec:exp test}.
The experiment is compared to numerical results.
A minimal model accounting for localized nonlinear losses is presented in Section~\ref{chap5:sec:model}.
Different simulation protocols are designed to reproduce the experiment (Section~\ref{chap5:sec:simprotocol}), to identify a set of RH dimensions that meets the playability criteria (Section~\ref{chap5:sec:simprotocol2}), and to study more specifically the global tuning of the second register (Section~\ref{chap5:sec:simprotocol3}).

\subsection{Experimental setup}\label{chap5:sec:exp}
\subsubsection{Clarinet-like resonators}\label{chap5:sec:exp:res}
\begin{figure}[H]
\centering
\includegraphics[width=.6\textwidth]{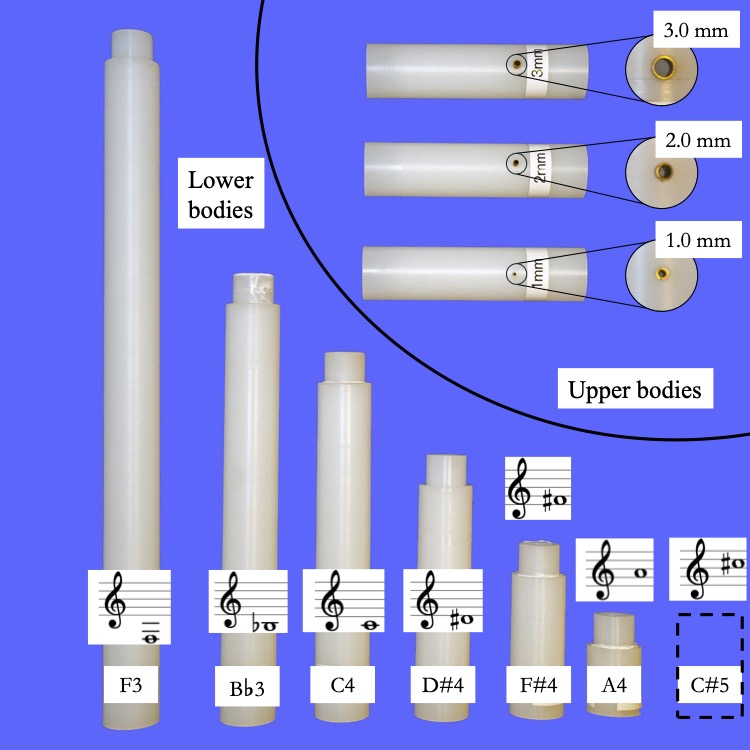}
\caption[Photography of the upper and lower bodies crafted for the experiment.]{Photography of the upper (labeled UB, at the top right corner) and lower bodies (labeled LB) crafted for the experiment.
Detailed views of the three register tubes.
}
\label{chap5:fig:methods:playTest:tubes}
\end{figure}

The resonators considered in this study are cylindrical tubes of length $L$ with one RH located at a constant distance $L_1=132$~mm from the tip of the mouthpiece, assuming that the mouthpiece has a length of 65~mm \cite{debut_analysis_2005} [Fig.~\ref{chap5:fig:0}(c)].
The resonators have an inner diameter $d=15.0$~mm  (or radius $R=d/2$).
Each resonator is made of one upper body and one lower body, represented in Figure~\ref{chap5:fig:methods:playTest:tubes}. 

Three different upper bodies are crafted, shown at the top right corner of Figure~\ref{chap5:fig:methods:playTest:tubes}.
Their length and inner diameter $d$ are the same.
A register tube is inserted at the same position for the three tubes. 
The register tube is made of brass, has a chimney height $L_h=10.0$~mm and an inner diameter $d_h=\{ 1.0, 2.0, 3.0 \} \pm 0.01$~mm.
In comparison, on a measured B$\flat$ clarinet (Buffet Crampon Festival), the register tube has an approximate length of $13.0$~mm and an inner diameter of $3.0$~mm on the outer side.
In the following, the upper bodies are labeled $\mathrm{UB}_x$ where $x$ is the corresponding diameter of the RH.

The lower bodies (LB) are six cylindrical tubes with the same inner diameter $d=15.0$~mm but different lengths. 
Seven different notes can then be played by connecting an upper body to one of the six lower bodies (playing notes F3, B$\flat$3, C4, D$\sharp$4, F$\sharp$4, A4), or to none (playing C$\sharp 5$).
The length $L_2$ from the RH to the open end of each lower body is given in Table~\ref{chap5:tab:lb}.
Note that switching the lower body changes the relative position $x_h$ of the RH with respect to the total length of the resonator, $x_h= L_1/(L_1+L_2)$. 

For each resonator  (one UB connected to one LB), the input impedance is measured when the hole is closed and opened (Fig.~\ref{fig:impedance}, Appendix~\ref{chap5:sec:Zin}).
An adaptation piece of length 65~mm is used to connect the resonator to the impedance sensor  \cite{dalmont2008new}.
The average first resonance frequency $f_1^c$ of the resonators  with the RH closed is written in Table~\ref{chap5:tab:lb}.
Note that when the hole is closed, discrepancies on $f_1^c$ due to different diameters of the RH are lower than 0.5~Hz.

\begin{table}[H]
\centering
\caption[Length $L_2$ from the register hole to the end of the lower bodies, for the seven lower bodies.]{Length $L_2$ from the register hole to the end of the lower bodies, for the seven lower bodies. Relative position $x_h=L_1/(L_1+L_2)$ of the register hole, with $L_1=132$~mm.
First resonance frequency $f_1^c$ for a closed register hole.}
\label{chap5:tab:lb}
\begin{tabular}{lccccccc}
\hline
Note name &  C$\sharp$5 & A4 & F$\sharp 4$ & D$\sharp 4$ & C4 & B$\flat 3$ & F3\\
$L_2$ (mm) & 40.0 & 75.6 & 116 & 166 & 223 & 268 & 403\\
$x_h$ (-) & 0.77 & 0.64 & 0.53 & 0.44 & 0.37 & 0.33 & 0.25\\
$f_1^c$ (Hz) & 509 & 398 & 333 & 278 & 234 & 208 & 153\\ \hline
\end{tabular}
\end{table}

\subsubsection{Playing tests protocol}\label{chap5:sec:exp test}
For each RH diameter, playing tests are carried out to assess how the probability of reaching the second register evolves for different notes.

\begin{figure}[H]
	\centering
	\includegraphics[width=\textwidth]{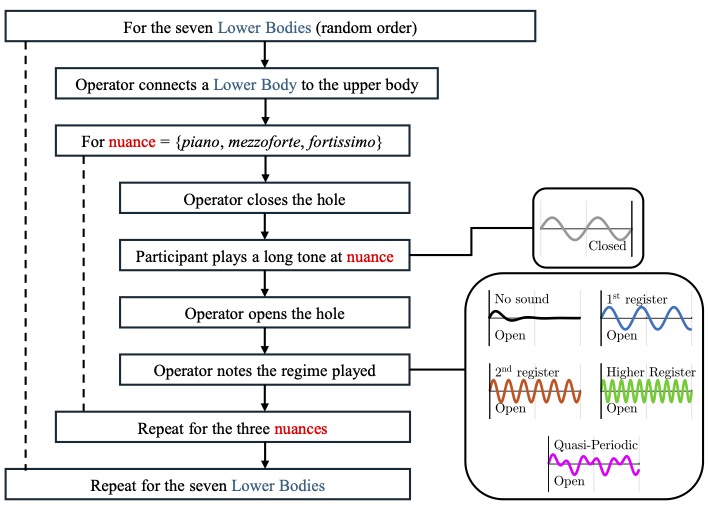}
	\caption[Summary scheme of a test Session.]{Summary scheme of a test Session. 
	For each Upper Body (or each diameter of the register hole), three sessions are completed.}
	\label{chap5:fig:schema_playtest}
\end{figure}

The test procedure incorporates some elements of the protocol from Szwarcberg \textit{et al.\ }(2024) \cite{szwarcberg2024second}.
It is described in Figure~\ref{chap5:fig:schema_playtest}. 
Participants are clarinetists of various levels: from very beginner to graduate. 
Each test focuses on a unique upper body.
Before starting the test, participant connects the mouthpiece (Buffet Crampon Urban Play) with the reed (Vandoren, strength 3.0) on the Upper Body. 
The upper body is fixed on a stand, so that the hands of the participant do not touch the instrument during the tests.

First, the participant seats, ready to play, blindfolded. 
The operator connects one of the lower bodies to the upper body and covers the RH with their finger. 
The participant then blows into the instrument to play ``long tone'', which is a first-register regime with a constant pressure and embouchure.
The participant is first asked to play \textit{piano} (softly).
While the musician blows, the operator opens the hole; the timing of the opening is announced pseudo-randomly in his headphones.
The resulting register or behavior is labeled following the code presented in Table~\ref{chap5:tab:playtest}.
This task is repeated for the nuance \textit{mezzo forte} (moderately loud) and \textit{fortissimo} (loud).
After these three long tones, the operator moves on to another lower body. 
The participant is asked again to perform three long tones with a gradually increasing blowing pressure.

During a test session, each of the seven lower bodies is blown three times: \textit{piano, mezzoforte, fortissimo}.
The order of the lower bodies is randomly selected.
Each participant completes three sessions.
The first is a training session whose results are not accounted for.
The complete test for one upper body last approximately 20~min.
Overall, for each Upper Body, $N=13$ participants participated to the tests.

\begin{table}[H]
\centering
\caption{Labels of the different behaviors observed after the opening of the hole in the experiment.
}
\label{chap5:tab:playtest}
\begin{tabular}{lccccc}
\hline
Label &  0 & 0.5 & 1 & 2 & 3\\
Regime produced & No sound & Quasi-periodic & 1\textsuperscript{st} Reg. & 2\textsuperscript{nd} Reg. & Higher Reg. \\
\hline
\end{tabular}
\end{table}

\subsection{Physical model}\label{chap5:sec:model}
This section presents the numerical model considered in this article.
The model was first introduced in Szwarcberg \textit{et al.} (2025)~\cite{szwarcberg2025register}.
It is recalled in this section for completeness.

Digital resonators are defined in sections~\ref{chap5:sec:model resonator}. 
From this geometry, the propagation of the acoustic waves in the resonator is described in sections \ref{chap5:sec: model eqs} and \ref{chap5:sec:bcs} through a delay lines formalism.
The method to account for nonlinear losses in the RH is described. 
The implementation for sound synthesis is presented in section \ref{chap5:sec:implementation}.

\begin{figure}[H]
	\centering
	\includegraphics[width=.8\textwidth]{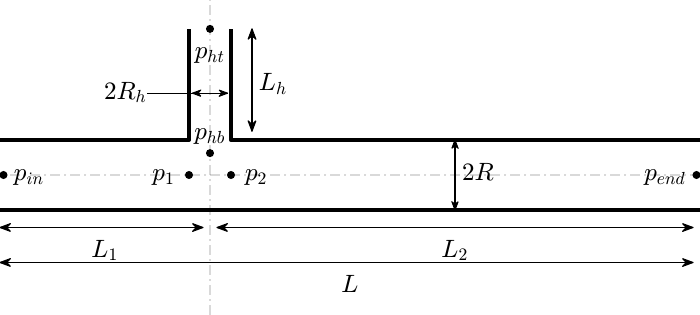}
	\caption{Definition of the digital resonators studied.}
	\label{chap5:fig:model:2}
\end{figure}

\subsubsection{Definition of the digital resonators}\label{chap5:sec:model resonator}

The digital resonator is presented on Figure \ref{chap5:fig:model:2}.
It is composed of a first tube of length $L_1=132$~mm, radius $R=7.5$~mm and cross-section $S=\pi R^2$.
The characteristic impedance of plane waves  propagating through the tube is $Z_c=\rho_0 c_0/S$ where $\rho_0=1.23$~kg$\cdot$m$^{-3}$ and $c_0=343~\mathrm{m\cdot s^{-1}}$.
The acoustic field in the first tube is described by the pressure at the left extremity $p_{in}$, and at the right extremity $p_1$.

The tube is branched to a side hole of length $L_h=10.0$~mm, radius $R_h=\{0.5,1.0,1.5 \}$~mm, cross-section $S_h=\pi R_h^2$, and characteristic impedance $Z_{ch}= \rho_0 c_0 / S_h$.
The acoustic field in the side hole is described by the pressure at the bottom of the hole $p_{hb}$ and at the top of the hole $p_{ht}$.

A second tube of variable length $L_2$ and cross-section $S$ is branched downstream from the side hole.
The acoustic field in this tube is described by the pressure at the left extremity $p_2$ and by the pressure at the right extremity $p_{end}$.

\subsubsection{Viscothermal losses}
Viscothermal losses are introduced through the complex wavenumber $\Gamma_i(s)$, where $s$ is the Laplace variable and $i=\{1,2,h \}$ refers to the index of the tube considered. 
The function $G_i(s)$ is defined, such that
\begin{align*}
G_i(s) &= e^{-\Gamma_i(s) L_i} = \lambda_i e^{-\epsilon_i \sqrt{s}} e^{-\tau_i s},
\end{align*}
with
\begin{align*}
\lambda_i &= e^{-({\alpha_2 \ell_v L_i})/{R_i^2}}, &
\epsilon_i &= \frac{\alpha_1 L_i}{R_i}\sqrt{\frac{2 \ell_v}{c_0}}, & \tau_i &= \frac{L_i}{c_0}, 
\end{align*}
where $\alpha_1=1.044$,  $\alpha_2=1.080$,  and $\ell_v= 4 \cdot 10^{-8}~\mathrm{m}$ [Chap.\ 5.5 of Chaigne and Kergomard (2016)] \cite{bible2016}.

In practice, $G_i(s)$ are approximated by a first-order low-pass filter and a delay $\tilde{G}_i(s)$, following the work from Guillemain \textit{et al.}\ (2005) \cite{guillemain2005real}.
Fractional delays $\tau_i$ are also taken into account by first-order filters proposed by Laakso \textit{et al}.\ (1996) \cite{laakso1996splitting}.

\subsubsection{Forward and backward-propagating pressure waves}\label{chap5:sec: model eqs}
In the following, time-domain variables are written in small letters ({e.g.}\ $p_2^+(t)$), and frequency-domain variables are written in capital letters ({e.g.}\ $P_2^+(s)$).

The propagation of the acoustic waves in the resonator is described through the forward and backward-propagating acoustic pressures $p^+$ and $p^-$.
They are related to the acoustic pressure and flow $(p, u)$ through the relationships:
\begin{align*}
p&= p^+ + p^-, & u &= \frac{p^+ - p^-}{Z},
\end{align*}
where $Z=Z_c$ in the main tube of cross-section $S$, and $Z= Z_{ch}$ in the side hole.

Since the tubes $L_1$, $L_2$ and $L_h$ are all cylindrical, the acoustic field can be described as transmission lines equations in the frequency domain, following Figure~\ref{chap5:fig:2}.
For the tube of length $L_1$:
\begin{align}
P_1^+ &= G_1 P_{in}^+, & P_{in}^- &= G_1 P_{1}^- .
\end{align}
For the tube of length $L_2$:
\begin{align}
P_{end}^+ &= G_2 P_{2}^+, & P_{2}^- &= G_2 P_{end}^- .
\end{align}
For the tube of length $L_h$:
\begin{align}
P_{ht}^+ &= G_h P_{hb}^+, & P_{hb}^- &= G_h P_{ht}^- .
\end{align}

\subsubsection{Boundary conditions}\label{chap5:sec:bcs}
The boundary conditions in the tube are described hereafter.

\paragraph{Radiation \\}
First, radiation from the open end is neglected: the pressure $p_{end}$ is written consequently as
\begin{equation}
 p_{end} = 0.
\end{equation}

\paragraph{Hole junction \\}
Secondly, since the RH has a small diameter and a long chimney length, the series impedances of the hole can be neglected [section 3.3.2.2 of Debut \textit{et al}.\ (2005)] \cite{debut_analysis_2005}.
The boundary conditions at the bottom of the hole are therefore given by:
\begin{align}
p_1 &= p_2, \label{chap5:eq:p1p2}  \\ 
p_2 &= p_{hb}, \label{chap5:eq:p2ph}  \\ 
u_1 &= u_2 + u_{hb}. \label{chap5:eq:flowHole}
\end{align}

\paragraph{Flow crossing the reed channel \\}
The next boundary condition involves $p_{in}$ and comes from the nonlinear characteristics of the flow entering the resonator. 
In this relationship, the acoustic flow $u_{in}$ depends on the difference between the blowing pressure of the musician $p_m$ and the pressure at the input of the instrument $p_{in}$.
By assuming that the jet experiences total turbulent dissipation  \cite{wilson_operating_1974} and modeling  the reed as a massless, undamped spring  \cite{ollivier2005idealized}, the nonlinear characteristics is defined as  \cite{dalmont2003nonlinear}:
\begin{equation}\label{chap5:eq:uin}
\hat{u}_{in} = \zeta [\hat{p}_{in} - \gamma +1]^+ \text{sgn}(\gamma - \hat{p}_{in}) \sqrt{|\gamma - \hat{p}_{in}|}, 
\end{equation}
where the function $[x]^+$ returns the positive-part of $x$, i.e.\ $ [x]^+ = (x + |x|)/2$.
The dimensionless blowing pressure is given by $\gamma= p_m/P_M$, where $P_M$ is the minimum pressure needed to close the reed channel in a quasi-static regime. 
Typical values of $P_M$ are in the range $P_M\in[4, 10]$~kPa \cite{dalmont2003nonlinear,atig2004saturation, dalmont_oscillation_2007, chatziioannou2012estimation, bergeot2014response,chabassier2022control}.
The parameter $\zeta$ represents the embouchure, with common values for the clarinet between 0.05 and 0.4  \cite{dalmont2003nonlinear, dalmont_oscillation_2007}.
The dimensionless quantities are defined as
\begin{align*}
\hat{p}_{in}&=p_{in} / P_M, & \hat{u}_{in} &=  u_{in}Z_c/ P_M.
\end{align*}

In Eq.\ \eqref{chap5:eq:uin}, the dynamics of the reed are neglected to obtain a direct relationship between $p^+_{in}$ and $p^-_{in}$. 
This relationship is given in Taillard \textit{et al}.\ (2010) \cite{taillard2010iterated} and is detailed in the Appendix of Bergeot \textit{et al.}\ (2014) \cite{bergeot2014response}.
It is expressed as:
\begin{align}\label{chap5:eq:Raman}
\hat{p}_{in}^+ = f_{\gamma \zeta}(\hat{p}_{in}^-)= \gamma - X[\gamma - 2 \hat{p}_{in}^-] - \hat{p}_{in}^-,
\end{align}
where the function $X$ is defined in Appendix A of Taillard \textit{et al}.\ (2010)  \cite{taillard2010iterated}.

\paragraph{Localized nonlinear losses in the register hole \\}\label{chap5:sec:model:Knl}
Localized nonlinear losses in the RH are modeled using the following boundary condition for $p_{ht}$:
\begin{equation}\label{chap5:eq:bcnl pv}
p_{ht}(t) = \rho_0 C_\text{nl} v_{ht}(t)|v_{ht}(t)|,
\end{equation}
where $v_{ht}$ is the acoustic speed at the top of the side hole, and $C_\text{nl}\geq 0$ is the nonlinear losses coefficient, which increases when the hole edges are sharper \cite{atig2004saturation}.
It is independent from the diameter of the hole \cite{dalmont_experimental_2002}.  
An explicit relationship between $p^+_{ht}$ and  $p^-_{ht}$ is given in Szwarcberg \textit{et al}.\ (2025) \cite{szwarcberg2025oscillation}:
\begin{equation}
p_{ht}^-(t) = r_\mathrm{nl}\left[p_{ht}^+(t)\right],
\end{equation}
where
\begin{align}\label{chap5:eq:bcnl}
r_\mathrm{nl}(x) = x \left(1- \frac{4}{1+ \sqrt{1+K_\mathrm{nl}|x|}} \right),
\end{align}
with $K_\mathrm{nl}=8C_\mathrm{nl}/(\rho_0 c_0^2)$.
For $K_\mathrm{nl}=0$, we get $r_\mathrm{nl}(x)$~$=$~$-x$, which corresponds to an open hole boundary condition.
As $K_\mathrm{nl} \to \infty$, $r_\mathrm{nl}(x)=x$, meaning the hole is closed.

In a dimensionless form, $r_\mathrm{nl}$ is rewritten as $\hat{p}_{ht}^-$~$=$~$\hat{r}_\mathrm{nl}\left[\hat{p}_{ht}^+\right]$, where
\begin{align}\label{chap5:eq:bcnl adim}
\hat{r}_\mathrm{nl}(x) = x \left(1- \frac{4}{1+ \sqrt{1+\hat{K}_\mathrm{nl}|x|}} \right),
\end{align}
with $\hat{K}_\mathrm{nl}= P_M K_\mathrm{nl}=0.2$, assuming a moderate reed closing pressure ($P_M=5.168$~kPa) \cite{dalmont2003nonlinear} and a hole with sharp edges \cite{atig2004saturation, dalmont_oscillation_2007}.

\subsection{Implementation of the physical model}\label{chap5:sec:implementation}
In this section, the synthesis algorithm is presented.
\subsubsection{Preliminary approximations}
\paragraph{Approximation for viscothermal losses \\}
To describe the complete system with a reasonable level of complexity, the filters $G_i(s)$ are approximated by a first-order low-pass filter and a delay, named $\tilde{G}_i(s)$,  for each of the three parts of the resonator:
\begin{equation}\label{chap5:eq:lowpass}
\tilde{G}_i(s) = \dfrac{b_i e^{-s \tau_i} }{1- a_i e^{-s/F_s}},
\end{equation}
where $F_s=44.1$~kHz is the sampling frequency. The sampling period is also defined by $T_s=1/F_s$.
Guillemain \textit{et al.\ }(2005) \cite{guillemain2005real} propose to compute $a_i$ and $b_i$ analytically (see Eqs.\ (15) and (16)) so that $|\tilde{G}_i|$ equals $|G_i|$ for two chosen frequencies.
Here, $|\tilde{G}_i|$ is fitted on the first and on the third resonance frequencies of the resonator.
This compromise is chosen to provide the same damping on the first mode as the full viscothermal losses model, and to prevent the over-damping of the fourth mode (which can be played experimentally) without favoring too much the second mode.
\paragraph{Approximation for fractional delays \\}
In the discrete-frequency domain, $\tilde{G}_i$ can be approximated by:
\begin{equation}
\tilde{G}_i(z) \simeq \frac{b_i z^{-D_i}}{1-a_i z^{-1}},
\end{equation}
where $z=e^{s/F_s}$ and $D_i = \lfloor F_s \tau_i \rfloor$ is the integer part of $F_s \tau_i$. 
For $c_0=343~$m/s and $F_s=44.1$~kHz, the minimal space step is $\Delta \ell = 7.8$~mm.
In this case, two RHs of length $L_h=7.8$~mm and $L_h=12.7$~mm would be indistinguishable, while they could lead to different results in practice.
To reduce this coarse discretization, Lagrange fractional delays filters of order 1 are employed, following Laakso \textit{et al.\ }(1996) \cite{laakso1996splitting}.
Filters $\tilde{G}_i$ are thus apprroximated by:
\begin{equation}
\tilde{G}_i(z) \simeq b_i\frac{ (1-\alpha_i)z^{-D_i} + \alpha_i z^{-(D_i+1)} }{1-a_i z^{-1}},
\end{equation}
where $\alpha_i = F_s \tau_i - D_i$ is the fractional part of $F_s \tau_i$.
\subsubsection{Synthesis algorithm}\label{chap5:app:synth}

\paragraph{Definition of the filtered variables \\}\label{chap5:app:var}
To simplify the following calculations, the following filtered and delayed variables are defined.
\begin{align*}
\Psi_1 &= P_{in}^+ \tilde{G}_1, & \Psi_2 &= P_{2}^+ \tilde{G}_2^2, \\
\Psi_{h1} &= P_{hb}^+ \tilde{G}_h, & \psi_{h2} &= r_\mathrm{nl}(\psi_{h1}), \\
 \Psi_{h3} &= \Psi_{h2} \tilde{G}_h,
\end{align*}
where the letter $\Psi$ refers to $z$-domain variables and $\psi$ refers to discrete-time variables.
\paragraph{Equation setting \\}\label{chap5:app:eq}
In this section, the boundary conditions presented in section \ref{chap5:sec:bcs} are combined to compute the acoustic field in the tube over time.
\subparagraph{Acoustic pressure in the side hole.\\
}
The equation of the conservation of the flow in the side hole (Eq.\ \eqref{chap5:eq:flowHole}) is developed:
\begin{align*}
p_2^+ - \psi_2 + \frac{S_h}{S}(p_{hb}^+ - \psi_{h3})= u_1.
\end{align*}
By substitution of $P_{in}^-$ from the equation $P_1=P_2$ (Eq.\ \eqref{chap5:eq:p1p2}) in the equation of the conservation of flow in the side hole, the following equation is obtained:
\begin{align*}
p_2^+ + \frac{S_h}{2S}(p_{hb}^+ - \psi_{h3}) - \psi_1= 0.
\end{align*}
In the equation obtained, $p_2^+$ is substituted by the equality $p_2=p_h$ (Eq.\ \eqref{chap5:eq:p2ph}) to write a causal relationship on $p_h^+$:
\begin{align}
p_{hb}^+ = \frac{2S}{2S + S_h} \left[\psi_1 - \psi_2 +\frac{S_h-2S}{2S}\psi_{h3} \right].\label{chap5:eq:phb+ causal}
\end{align}

\subparagraph{Progressive pressure in the second tube.\\
}
After computing $p_{hb}^+$ through Eq.\ \eqref{chap5:eq:phb+ causal}, $p_{hb}^+$ is computed from the equality $p_2=p_h$ (Eq.\ \eqref{chap5:eq:p2ph}):
\begin{align}
p_2^+ = p_{hb}^+ + \psi_2 + \psi_{h3}.
\label{chap5:eq:p2+ causal}
\end{align}

\subparagraph{Pressure at the input of the resonator.\\
}
The regressive pressure at the input $p_{in}^-$ is computed after filtering the equality $p_1=p_2$ (Eq.\ \eqref{chap5:eq:p1p2}) by $\tilde{G}_1$:
\begin{align}
p_{in}^- = \mathcal{Z}^{-1} \left[ \tilde{G}_1 \cdot \left( P_2^+ - \Psi_2 - \Psi_1 \right) \right]. \label{chap5:eq:pinm causal}
\end{align}
The progressive pressure at the input $p_{in}^+$ is finally deduced through the equation of the nonlinear characteristic of the flow crossing the reed channel (Eq.\ \eqref{chap5:eq:Raman}):
\begin{align}
p_{in}^+ / P_M= f_{\gamma \zeta}(p_{in}^- / P_M),
\label{chap5:eq:pinp causal}
\end{align}
where $P_M$ is the reed closing pressure defined in section \ref{chap5:sec:bcs}.

\paragraph{Practical implementation \\}

\begin{figure}[h]
	\centering
	\includegraphics[width=\textwidth]{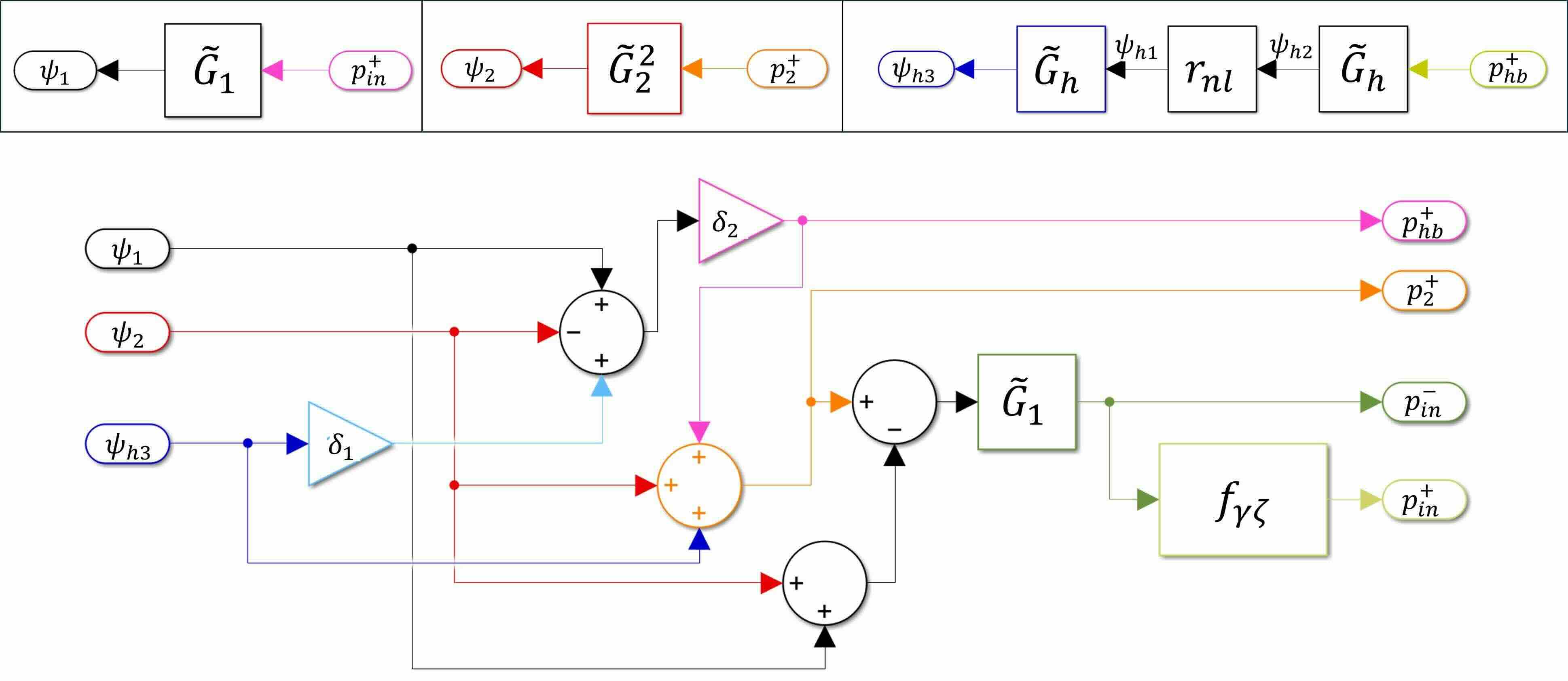}
	\caption[Synthesis scheme of the resonator. ]{Synthesis scheme of the resonator. The gain $\delta_1$ multiplying $\psi_{h3}$ is $\delta_1=(S_h-2S)/(2S)$. 
	The gain $\delta_2$ outputting $p_{hb}^+$ is $\delta_2=2S/(2S+S_h)$.}
	\label{chap5:fig:model:block}
\end{figure}

The synthesis scheme is represented on Figure \ref{chap5:fig:model:block}.
Initial conditions for ${p}_{in}^+$, ${p}_{in}^-$, ${p}_{hb}^+$ and $p_2^+$ are all set to zero.
At each time step:
\begin{enumerate}
\item Update the filtered acoustic variables $\psi_1, \psi_2, \psi_{h1}, \psi_{h2}, \psi_{h3}$, defined in section \ref{chap5:app:var} (see also the three frames at the top of Figure \ref{chap5:fig:model:block}).
In practice, these variables are computed through the \texttt{filter} Matlab function \cite{matlab_filter_function}.
\item Compute $p_h^+$ from Eq.\ \eqref{chap5:eq:phb+ causal},  followed by $p_2^+$ from Eq.\ \eqref{chap5:eq:p2+ causal}, then $p_{in}^-$ from Eq.\ \eqref{chap5:eq:pinm causal}, and finally $p_{in}^+$ from Eq.\ \eqref{chap5:eq:pinp causal}.
\end{enumerate}
These two steps are repeated for each next temporal iteration.

In all simulations, the synthesis model is implemented in a dimensionless formalism, through the use of variables $\hat{p}_{in}^+$, $\hat{p}_{in}^-$, $\hat{p}_{hb}^+$, and $\hat{p}_{2}^+$.
This only involves to replace the equation of $\psi_{h2}$ (see section \ref{chap5:app:var}) by $$ \hat{\psi}_{h2} = \hat{r}_\mathrm{nl}(\hat{\psi}_{h1}).$$

\subsection{Simulation protocols}
\subsubsection{Simulation protocol 1: Reproduction of the experiment}\label{chap5:sec:simprotocol}

This section details the protocol leading to the results of Section~\ref{chap5:sec:results:1}.
From the model presented in section \ref{chap5:sec:implementation}, time-domain simulations are carried out to reproduce the experiment. 
Different resonator geometries are considered (see section~\ref{chap5:sec:simu:var_geom}).
For each geometry, simulations are realized for given set of dimensionless blowing pressure $\gamma$ and embouchure parameter $\zeta$.

\begin{figure}[H]
\centering
\includegraphics[width=\textwidth]{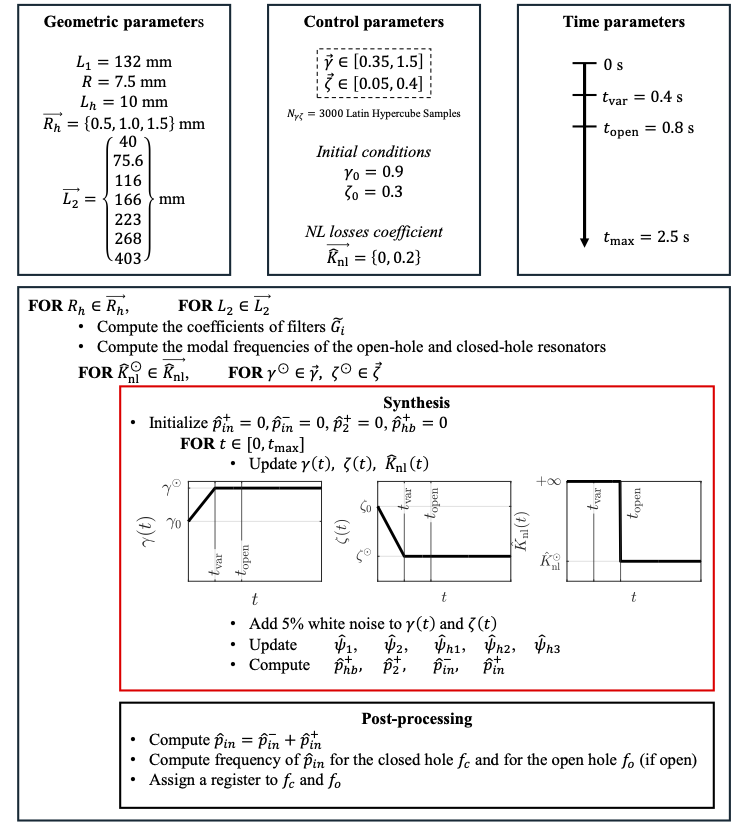}
\caption{Simulation protocol for the reproduction of the playing tests (Figure \ref{chap5:fig:1}).}
\label{chap5:fig:sim:protocol}
\end{figure}

Each simulation lasts for $t_\mathrm{max}=2.5$~s.
At the beginning, the hole is closed ($\hat{K}_\mathrm{nl}=+\infty \Leftrightarrow \hat{r}_\mathrm{nl}(x)=x$).
Around $t_\mathrm{open}=0.8$~s, the hole is instantaneously opened by changing $\hat{K}_\mathrm{nl} $ to the value chosen for the simulation.
At the end of each simulation, the frequency of the input pressure signal is measured slightly before the hole opens, and at the end of the simulation.
The simulation protocol is summarized in Figure \ref{chap5:fig:sim:protocol}.

\paragraph{Variation of the geometry of the resonator \\}\label{chap5:sec:simu:var_geom}

The resonators considered all have the same geometry as in the experiment.
All values are given in the ``Geometric parameters'' block of Figure~\ref{chap5:fig:sim:protocol}.




\paragraph{Variation of the control parameters \\}\label{chap5:sec:simu:var_params}
Each simulation is run with a set of control parameters $\gamma^\odot$ and $\zeta^\odot$. 
The control parameters space $(\gamma, \zeta)$ is mapped by $N_{\gamma \zeta}=3000$ Latin hypercube samples (each sample is alone in each axis-aligned
hyperplane containing it  \cite{mckay2000comparison}).
They are distributed in the range $\gamma \in [0.35, 1.5]$ and $\zeta \in [0.05, 0.4]$.
We verified that $N_{\gamma \zeta}=3000$ enables to estimate the proportion of each register in this range of control parameters with an error lower than 5\%.

\begin{figure}[H]
	\centering
	\includegraphics[width=.8\textwidth]{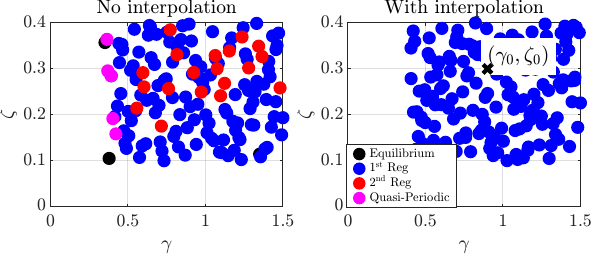}
	\caption[Role of the interpolation of $(\gamma,\zeta)$ from starting values $(\gamma_0, \zeta_0)$ in the ability to produce the first register for the closed hole is closed.]{Role of the interpolation of $(\gamma,\zeta)$ from starting values $(\gamma_0, \zeta_0)$ in the ability to produce the first register for the closed hole is closed.
	On the left-hand side, simulations are directly run to the target values $(\gamma^\odot, \zeta^\odot)$, without interpolation. Different regimes are produced.
	On the right-hand side, simulations start at $(\gamma_0, \zeta_0)$;  control parameters are then interpolated to the target values.
	The first register is the only regime produced. }
	\label{chap5:fig:methods:interpolation}
\end{figure}

When the hole is closed, to enable the model to play a stable register for a target blowing pressure and embouchure $(\gamma^\odot, \zeta^\odot )$, the control parameters are first linearly interpolated from $(\gamma_0, \zeta_0)=(0.9,0.3)$ to $(\gamma^\odot, \zeta^\odot)$ for a duration $t_\mathrm{var}=0.4$~s.
Preliminary tests confirm that these values are suitable starting points, as shown on Figure \ref{chap5:fig:methods:interpolation}.
From time $t>t_\mathrm{var}$, the control parameters are kept at their target values until the end of the simulation at time $t_\mathrm{max}=2.5$~s.
The register played is then determined.

To help the unstable regimes to lose their stability after opening the hole, 5~\% of white noise is added to $\gamma(t)$ and $\zeta(t)$ \cite{szwarcberg2024second}.

\paragraph{Opening of the hole: variation of the nonlinear losses coefficient \\}\label{chap5:sec:varKnl}

\begin{figure}[H]
	\centering
	\includegraphics[width=.8\textwidth]{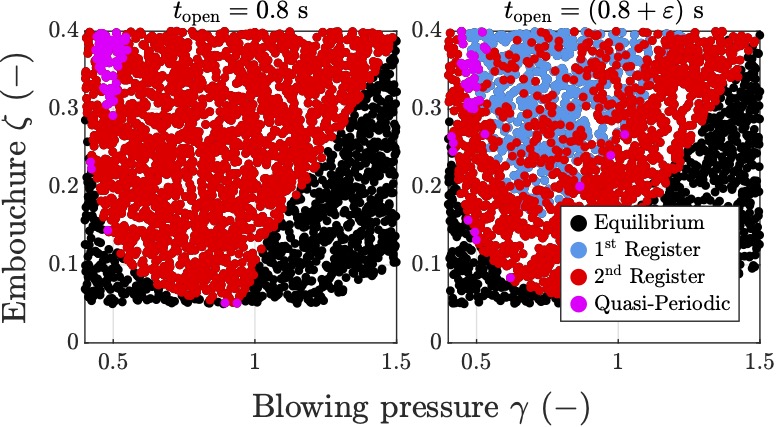}
	\caption[Evidence of phase-induced tipping when the register hole is opened.]{Evidence of phase-induced tipping when the register hole is opened. 
	When the hole is always opened at the same time (left panel), simulation results seem to highlight that the first register is unstable.
	However, when the hole is opened at a random time (right panel), a whole region is discovered in which the first and second register are both stable.
	Simulation parameters: $\hat K_\mathrm{nl}=0, d_h=3$~mm, $R=$, $L_1=132$~mm, $L_2=913$~mm, $R=7.5$~mm.  }
	\label{chap5:fig:methods:phasetipping}
\end{figure}

To account for the possible variability induced by phase-tipping phenomena \cite{alkhayuon2021phase, szwarcberg2025register}, the timing of the hole opening is randomly varied for each simulation:
$$ t_\mathrm{open} = 0.8 + \varepsilon, \quad  -T_1^c<\varepsilon< T_1^c, $$
where $T_1^c=1/f_1^c$ is the period of the first mode of the resonator with the hole closed. 
The modal frequencies are computed through the Transfer Matrix Method.
The transfer matrices account for the approximations $\tilde{G}_i$ on filters $G_i$.
An evidence of phase-induced tipping is shown in Figure \ref{chap5:fig:methods:phasetipping}.
 
 At $t=t_\mathrm{open}$, the hole is opened and the nonlinear losses coefficient switches from $\hat{K}_\mathrm{nl}=+\infty$ to $\hat K_\mathrm{nl}=0$ (nonlinear losses are ignored) or $\hat K_\mathrm{nl}=0.2$ (sharp hole edges, medium reed stiffness and mouthpiece aperture).

\paragraph{Post-processing: frequency and register classification \\}

\begin{figure}[H]
	\centering
	\includegraphics[width=\textwidth]{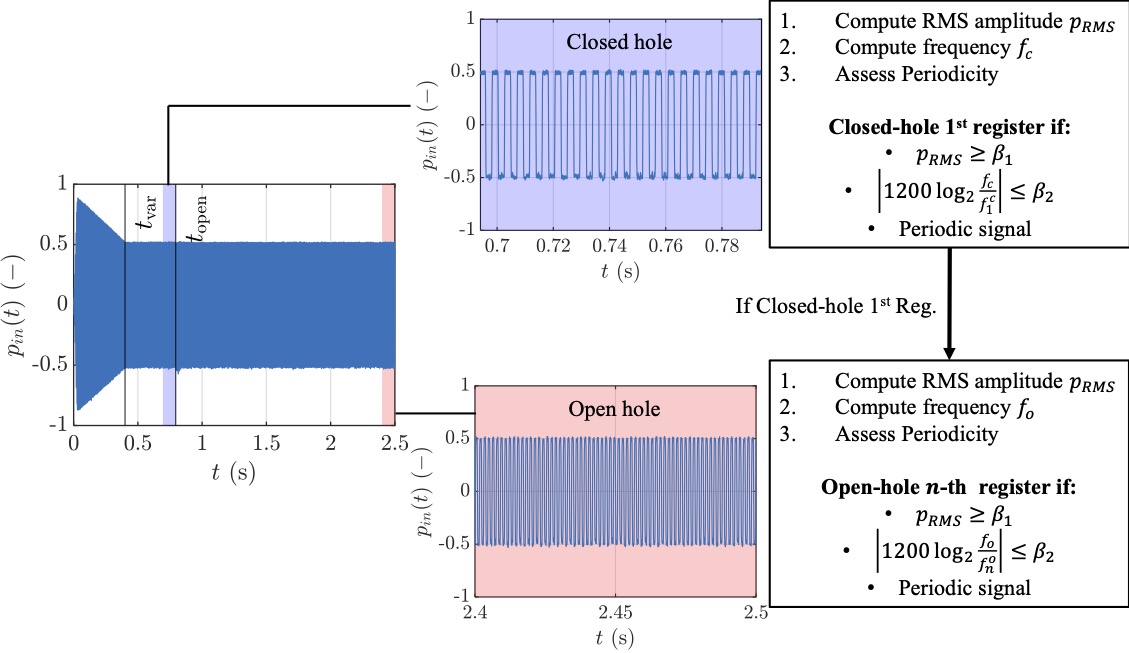}
	\caption[Scheme of the Register assignment process.]{Scheme of the Register assignment process. The frequency is computed from the zero-crossing of the signal.
	Periodicity is assessed by thresholding the variance of the zero-crossing rate of the signal. 
	$\beta_1=0.01$, $\beta_2=200$~cents. }
	\label{chap5:fig:methods:regs}
\end{figure}

After each simulation, the acoustic pressure at the input, $\hat{p}_{in} = \hat{p}_{in}^- + \hat{p}_{in}^+$, is analysed in two time intervals:  
just before the hole opens, $t \in [t_\mathrm{open} - 0.1, t_\mathrm{open}]$, and at the end of the simulation, $t \in [t_\mathrm{max} - 0.1, t_\mathrm{max}]$.
In both intervals, the frequency is measured and the register is identified.  
The process is illustrated in Figure~\ref{chap5:fig:methods:regs}.

A signal is said to belong to the $n$-th register if:
\begin{itemize}
	\item  its frequency is within $\beta_2 = 200$ cents of the $n$-th modal frequency of the resonator;
	\item  its amplitude is sufficiently high ($p_\mathrm{RMS} \geq \beta_1$  with $\beta_1 = 0.01$);
	\item the signal is periodic.
	Quasi-periodicity is assessed using a criterion based on the variation in zero-crossing intervals.
\end{itemize}

In the results, we only retain simulations that produce the first register before the hole opens, in agreement with the experimental observations.

\subsubsection{Simulation protocol 2: Design parameters for an efficient register hole}\label{chap5:sec:simprotocol2}

This section describes the procedure leading to the results presented in Figure~\ref{chap5:fig:2}.
On the clarinet, opening the RH should allow the production of the second register for all notes from E3$\to$B4 to F4$\to$C6.  
In addition, to ensure that the instrument is fully chromatic, opening the RH while playing A4 must raise the pitch by a semitone, i.e., to \textit{throat} B$\flat$4.  
The goal is to determine how the playability of the second register and of the B$\flat$4 depends on the hole’s diameter $d_h$, chimney length $L_h$, and position $L_1$.

\paragraph{Geometric setup\\}

We sample the 3D space of RH parameters $(d_h, L_1, L_h)$ as follows:
\begin{itemize}
	\item $d_h$ ranges from 1.0~mm to 4.0~mm in 0.5~mm increments
	\item $L_1$ ranges from 70~mm to 275~mm in 5~mm increments
	\item $L_h$ ranges from 5~mm to 19~mm in 2~mm increments
\end{itemize}

To address the requirements stated above, three total resonator lengths are studied:
\begin{center}
	E3: $L = 558$~mm; \hspace{1cm} F4: $L = 263$~mm; \hspace{1cm} A4: $L = 208$~mm.
\end{center}

For the E3 and F4 configurations, we test whether the RH can reliably produce the second register.  
For the A4 configuration, we test whether the RH can reliably produce the first register and raise the pitch by a semitone.

\paragraph{Cartography \\}

For each combination of $d_h$, $L_1$, $L_h$, and $L$ listed above, we compute a complete map of RH behavior in the $(\gamma, \zeta)$ control parameter space.  
The simulation setup follows the same protocol as in Section~\ref{chap5:sec:simprotocol}.  
The parameters are: $N_{\gamma \zeta} = 3000$, $\hat{K}_\mathrm{nl} = 0.2$, $t_\mathrm{var} = 0.4$~s, $t_\mathrm{open} = 0.8$~s, and $t_\mathrm{max} = 2.5$~s.  
Each simulation (i.e., one set of $(\gamma,\zeta)$ values) is classified into a register based on the acoustic amplitude, signal periodicity, and the playing frequency ratio between the open and closed hole configurations (Figure~\ref{chap5:fig:methods:regs}).

\paragraph{Post-processing \\}

For each cartography, we compute the proportion of regimes obtained after the hole is opened.

\begin{itemize}
	\item \textbf{For E3 and F4:}\\
	A RH configuration is considered playable if more than 50\% of simulations lead to the second register.  
	It is considered to play in tune if, among the $N_{R2}$ simulations producing the second register, the average frequency ratio between the open and closed hole corresponds to a musical twelfth, with a tolerance of $\pm5$~cents:
	\begin{equation}\label{chap5:eq:inharmNum}
	-5 \leq 1200 \, \mathrm{log}_2 \left[ \frac{1}{N_{R2}}\sum_i^{N_{R2}} \frac{(f_\mathrm{play}^\mathrm{open})_i}{3(f_\mathrm{play}^\mathrm{closed})_i} \right] \leq 5
	\end{equation}

	\item \textbf{For A4:}\\
	Two conditions must be satisfied for a configuration to be considered playable for the \textit{throat} B$\flat$4:
	\begin{enumerate}
		\item More than 50\% of the simulations must result in the first register.
		\item Among the $N_{R1}$ simulations that produce the first register, the average frequency shift between the open and closed hole must be close to a semitone.\\
		Since clarinetists often lower the pitch of the \textit{throat} B$\flat$4 by covering additional tone holes, we allow some tolerance.  
		We set the acceptable pitch range between $-5$~cents and $+25$~cents relative to a semitone.  
		The playability condition is:
		\begin{equation}
		95 \leq 1200 \, \mathrm{log}_2 \left[ \frac{1}{N_{R1}}\sum_i^{N_{R1}} \frac{(f_\mathrm{play}^\mathrm{open})_i}{(f_\mathrm{play}^\mathrm{closed})_i} \right] \leq 125
		\end{equation}
	\end{enumerate}
	\textit{Throat} B$\flat$4 is considered to be in tune if the pitch shift falls within a semitone, with a tolerance of $\pm5$~cents:
	\begin{equation}\label{chap5:eq:tuneBflat}
	95 \leq 1200 \, \mathrm{log}_2 \left[ \frac{1}{N_{R1}}\sum_i^{N_{R1}} \frac{(f_\mathrm{play}^\mathrm{open})_i}{(f_\mathrm{play}^\mathrm{closed})_i} \right] \leq 105
	\end{equation}
\end{itemize}

Finally, the register proportions and average playing frequency ratios are interpolated over the $(d_h, L_1, L_h)$ space.  
The resulting maps are presented in Figure~\ref{chap5:fig:2}.

\subsubsection{Simulation protocol 3: Global tuning of the second register}\label{chap5:sec:simprotocol3}
\begin{figure}[H]
	\centering
	\includegraphics[width=.9\textwidth]{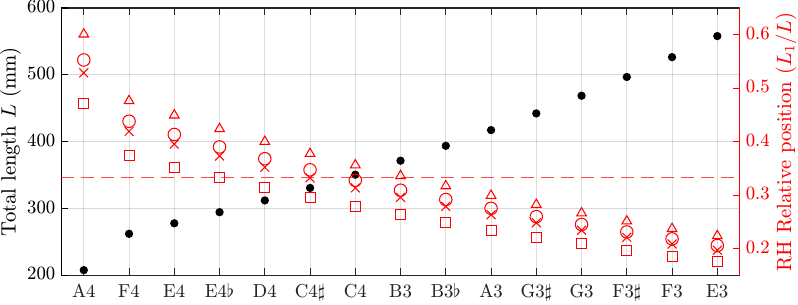}
	\caption[Total resonator length corresponding to each simulated note, and relative register hole position.]{Total resonator length $L$ corresponding to each simulated note, and relative register hole position ($L_1/L$), for the configurations of Table~\ref{chap5:tab:protocol:3}. 
	The red dashed line shows the pressure node of the second mode shape, $L_1 = L/3$.}
	\label{chap5:fig:protocol:3:length}
\end{figure}

This section describes the procedure leading to the results of Figure~\ref{chap5:fig:3}.
Four different RH configurations $(d_h, L_1, L_h)$ are considered, detailed in Table \ref{chap5:tab:protocol:3}.
For each configuration, hole openings simulations are carried out for all semitones between E3 and F4, as well as the A4.  (15 different notes).
Notes F$\sharp$4, G4 and G$\sharp$4 are excluded, since they are not typically used with an open RH.
The corresponding lengths are given in Figure~\ref{chap5:fig:protocol:3:length}.

For each RH configuration and each note, playability maps are realized with the same simulation parameters as the previous case (Section~\ref{chap5:sec:simprotocol2}).
After assessing the playability of each note, the average playing frequency is computed.

\begin{table}[H]
	\centering
	\caption{Register hole parameters used in the simulations of Section~\ref{chap5:sec:results:3}.}
	\label{chap5:tab:protocol:3}
	\begin{tabular}{cccc}
	\hline
	Label & $d_h$ (mm) & $L_1$ (mm) & $L_h$ (mm) \\	
	\hline
	$\bigtriangleup$ & 3.0 & 125 & 7.0 \\
	 $\bigcirc$ & 3.0 & 115 & 13\\
	 $\square$ & 3.0 & 98 & 20 \\
	 $\times$ & 2.0 & 110 & 13\\
	 \hline
	\end{tabular}
\end{table}

\section{Results}
\subsection{Nonlinear losses in the register hole govern register transitions}\label{chap5:sec:results:1}

We investigate how the diameter of the RH affects the production of the second register.  
Seven notes and three hole diameters are tested experimentally, through a playing test procedure detailed in Section~\ref{chap5:sec:exp}.  
Simulations are used to interpret the underlying mechanisms, with and without nonlinear losses.
We define the physical model in Section~\ref{chap5:sec:model} and detail its implementation in Section~\ref{chap5:sec:implementation}.
The numerical protocol is presented in Section~\ref{chap5:sec:simprotocol}.
Methods and data are accessible online \cite{szwarcberg2025data}.

\begin{figure}[H]
 	\centering	
 	\includegraphics[width=\textwidth]{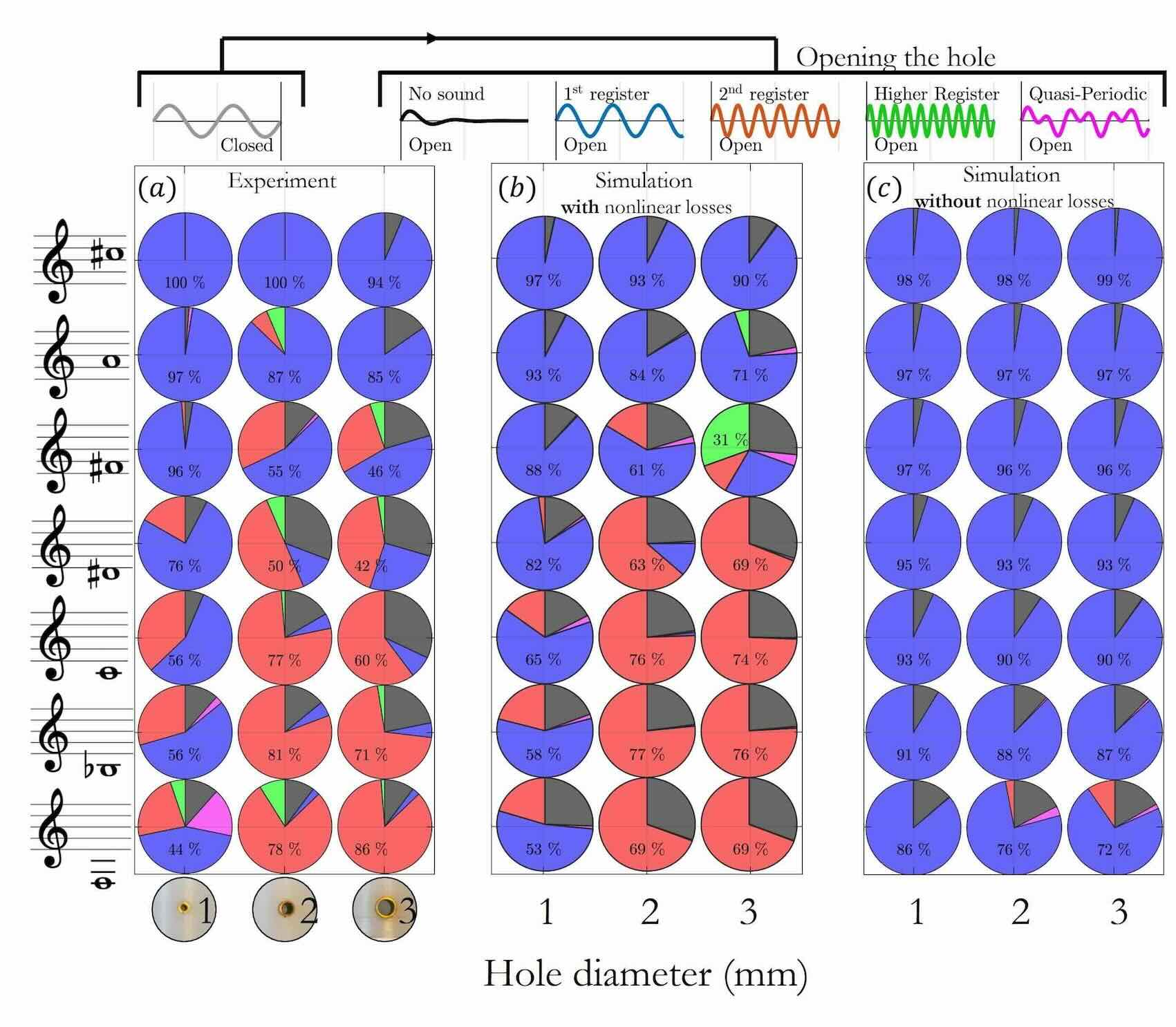}
 	\caption[Experimental and simulation results: register production distribution for different notes and register hole diameters.]{Experimental and simulation results: register production distribution for different notes and register hole diameters. 
(a): experimental results. 
(b): simulation with localized nonlinear losses in the register hole $(\hat{K}_\mathrm{nl}=0.2)$.
(c): simulation without localized nonlinear losses in the register hole ($\hat{K}_\mathrm{nl}=0$).
For each pie chart, the percentage of the most represented register is specified.
Each color refers to a type of regime obtained after opening the register hole.
Corresponding regimes are indicated at the top of the figure.
Statistical investigations are provided in Appendix~\ref{chap5:sec:stats}.
}
\label{chap5:fig:1}
 \end{figure}

Experimental results are shown in Figure~\ref{chap5:fig:1}(a).  
For each note and hole diameter, a pie chart indicates the distribution of playing regimes following the opening of the RH.  
Here, all nuances (\textit{piano, mezzoforte, fortissimo}) are aggregated.
The second register is shown in red.  
Statistical analyses supporting these results are provided in the Supplementary Data (Sec.~\ref{chap5:sec:stats}).

For the two highest notes (C$\sharp$5 and A4), the first register dominates.  
This outcome is consistent with the measured input impedance curves (Fig.~\ref{fig:impedance}, Sec.~\ref{chap5:sec:Zin}), where opening the RH produces only a minor shift in the resonances.

For the five lowest notes, 2.0~mm and 3.0~mm holes favor the second register significantly more than the 1.0~mm hole (Fig.~\ref{fig:stats1}, Sec.~\ref{chap5:sec:stats}).  
The 1.0~mm hole mainly produces the first register for all notes.  
It is also the only configuration in which playing \textit{fortissimo} increases the occurrence of the first register compared to \textit{piano} [Fig.~\ref{fig:stats2}(b), Sec.~\ref{chap5:sec:stats}].  
These differences are explained by the presence of localized nonlinear losses in small-diameter holes.  
As the acoustic velocity is higher in narrower holes, the nonlinear resistance accounting for vortex shedding becomes significant.  
The RH then behaves acoustically as if it were closed \cite{debut_analysis_2005}, favoring the first register.

No significant difference is experimentally observed between 2.0~mm and 3.0~mm holes in their ability to produce the second register (Fig.~\ref{fig:stats1}, Sec.~\ref{chap5:sec:stats}).

Simulation results are shown in Figures~\ref{chap5:fig:1}(b) and \ref{chap5:fig:1}(c).  
When localized nonlinear losses are neglected (Figure~\ref{chap5:fig:1}(c)), the model fails to produce the second register.  
The first register persists, indicating insufficient damping of the first mode after the hole opens.

Introducing localized nonlinear losses into the model enables the production of the second register (Figure~\ref{chap5:fig:1}(b)).  
The model captures the key trends observed in the experiment.
The 1~mm hole behaves distinctly from the larger holes, producing mainly the first register for all notes.  
However, the second register remains under-represented in simulation compared to the experiment.

Figure~\ref{fig:supp:sim:nuance} (Appendix~\ref{chap5:fig:supp:sim}) clarifies this discrepancy.  
It shows that the second register in experiments appears mostly for soft dynamics, consistent with the ``closed-hole effect'' due to nonlinear losses  \cite{debut_analysis_2005,bible2016}.  
In simulations, however, low-pressure signals are underrepresented, which likely accounts for the mismatch.

Simulations for 2.0~mm and 3.0~mm holes reproduce the observed trends: the second register is mainly produced for the four lowest notes.
One notable difference appears on Figure~\ref{chap5:fig:1} for F$\sharp$4 with the 3.0~mm hole.  
Simulations highlight a dominant production of the third register, while this occurred only twice experimentally.  
This is due to the register hole’s proximity to the second pressure node of the third mode shape ($3L/5$), which promotes the excitation of that mode.  
Figure~\ref{fig:supp:gmax} (Appendix~\ref{chap5:fig:supp:sim}) suggests that very high blowing pressures could favor the production of the third register.  
This indicates that high-amplitude regimes tend to be overrepresented in the simulations.  
The same bias likely explains the overproduction of the equilibrium (no-sound) regime for the note F3.

In summary, a sparse clarinet model accounting for localized nonlinear losses in the RH reproduces the register dynamics observed experimentally.  
Nonlinear losses are essential to explain the appearance of the second register.  
This model enables systematic investigation of how RH geometry affects both playability and tuning.

\subsection{The fragile balance behind the second register of the clarinet}\label{chap5:sec:results:2}

\begin{figure}[H]
	\centering
	\includegraphics[width=1\textwidth]{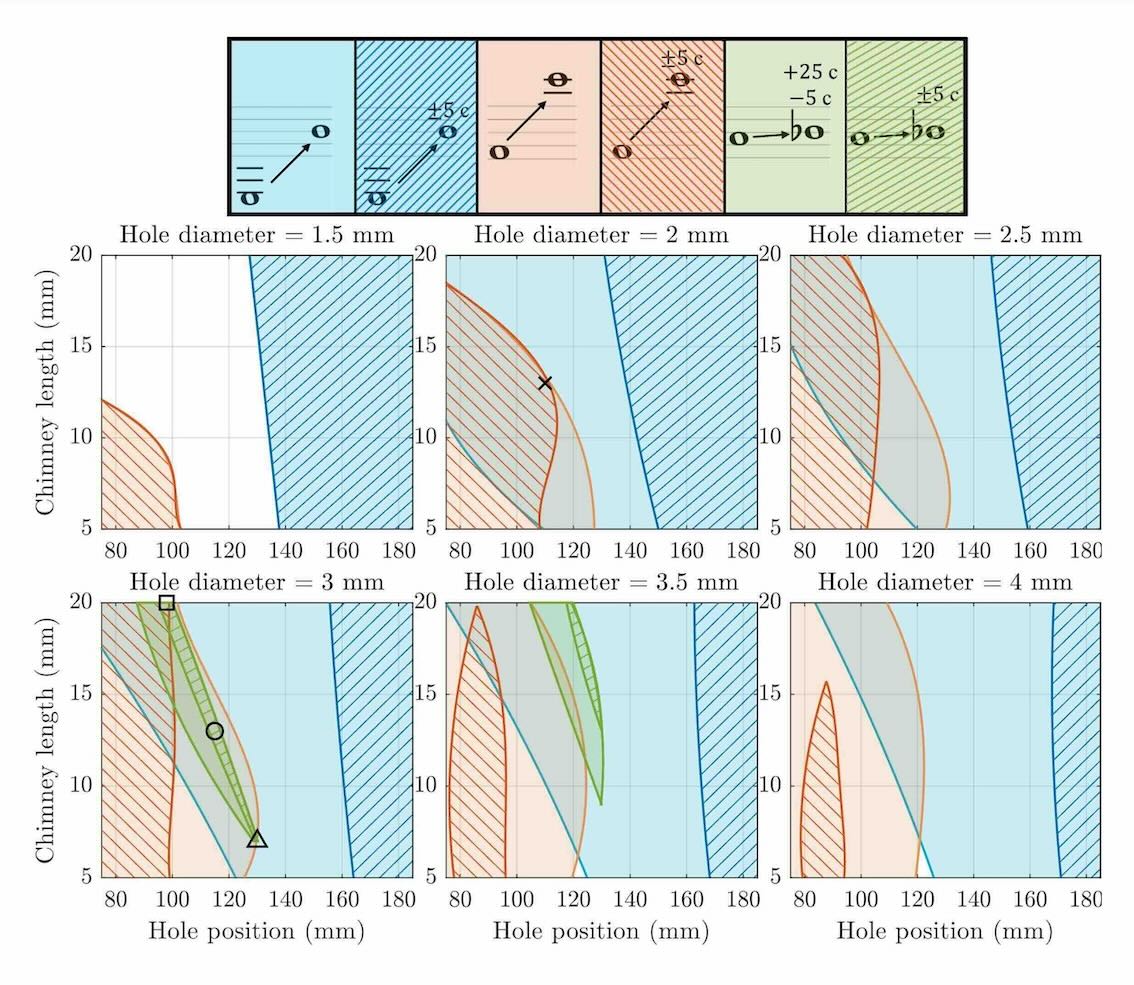}
	\caption[Playability and tuning maps for the second register and the \textit{throat} B$\flat$4, as a function of register hole position, diameter, and chimney length.]{Playability and tuning maps for the second register and the \textit{throat} B$\flat$4, as a function of register hole position, diameter, and chimney length.
Orange and blue regions indicate that the second register is reliably produced after opening the register hole.  
Orange corresponds to the highest note (F4$\to$C6), and blue to the lowest (E3$\to$B4).  
Hatchings show configurations where the second register is in tune [see Eq.~\eqref{chap5:eq:inharmNum}].
Green regions indicate register hole parameters that allow raising the pitch from A4 to B$\flat$4.  
Green hatching shows where the B$\flat$4 is in tune [Eq.~\eqref{chap5:eq:tuneBflat}].
The corresponding resonator lengths are $L = \{558, 263, 208\}$~mm for E3$\to$B4, F4$\to$C6, and A4$\to$B$\flat$4, respectively.
The four markers highlight register hole parameters used in Figure~\ref{chap5:fig:3}.  
The simulation protocol is described in Section~\ref{chap5:sec:simprotocol2}.
	}
	\label{chap5:fig:2}
\end{figure}

As presented in the Introduction, for a \textit{playable} instrument over the first and second registers, the design of the RH should satisfy three criteria:
\begin{enumerate}
	\item \textit{overblow} from the E3 to the F4;
	\item minimize \textit{inharmonicity} between the first and the second registers;
	\item produce the \textit{throat} B$\flat$4 when opening the RH from the A4.
\end{enumerate}
All three conditions should be satisfied over a wide range of blowing pressures and embouchure configurations.
These characteristics are explored following the protocol presented in Section~\ref{chap5:sec:simprotocol2}.
Results are presented in Figure~\ref{chap5:fig:2}.

The regions where the two extreme notes of the second register (E3$\to$B4 and F4$\to$C6) can be reliably produced appear where both orange and blue color patches overlap.  
This overlap only exists when the RH diameter is strictly greater than 1.5~mm, and lower than 4.0~mm.  
Within this range, suitable combinations of position and chimney length allow the production of the second register both for the lowest (E3$\to$B4) and the highest notes (F4$\to$C6).
As the diameter increases, the overlapping region shrinks.  

The regions of the design space minimizing inharmonicity are represented in hatchings. 
Even within the regions where the E3 and the F4 can both overblow, perfect tuning is never achieved for both notes.

Finally, the production of the \textit{throat} B$\flat$4 imposes a severe constraint.  
Only a 3.0 mm diameter allows both accurate tuning of this note and reliable second register production.  
This compromise is achieved within a narrow band of chimney lengths and positions.

\subsection{Relaxing constraints to improve tuning}\label{chap5:sec:results:3}

The previous section shows that producing the \textit{throat} B$\flat$4 by opening the RH from A4 imposes a strong geometric constraint. 
In particular, it requires a RH diameter around 3.0~mm, strictly greater than 2.5~mm and lower than 3.5~mm. 
All viable configurations lie within a narrow region in the parameter space.

\begin{figure}[H]
	\centering
	\includegraphics[width=\textwidth]{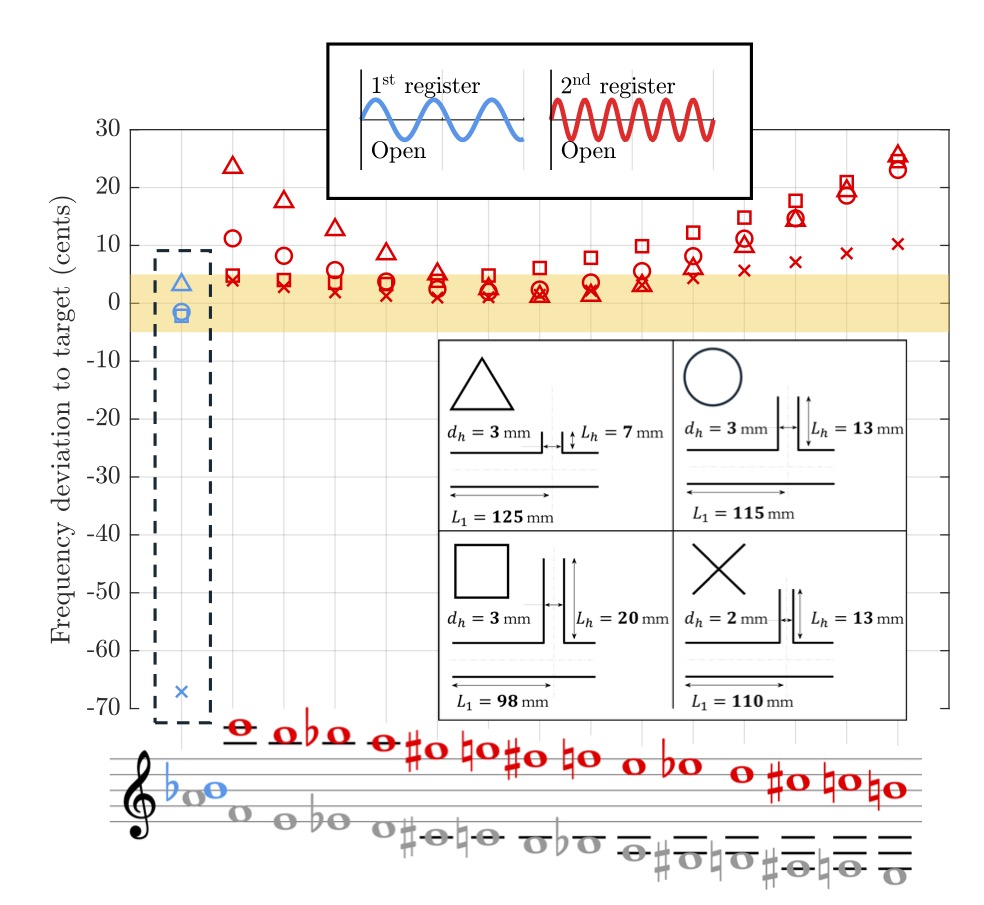}
	\caption[Tuning of the \textit{throat} B$\flat$4 and all notes of the second register, for four register hole configurations.]{Tuning of the \textit{throat} B$\flat$4 and all notes of the second register, for four register hole configurations ($\bigtriangleup, \bigcirc, \square, \times$). 
	Markers color refers to the type of register obtained when opening the hole.
	A marker located in the yellow band ($\pm5$ cents) means that the note played by the model is in tune.
	The register hole characteristics corresponding to each marker are specified on the legend.
	These configurations are also represented on Figure \ref{chap5:fig:2}.
	The simulation protocol is described in Section~\ref{chap5:sec:simprotocol3}.
	}
	\label{chap5:fig:3}
\end{figure}

Figure~\ref{chap5:fig:3} shows how tuning varies within this region. 
With a 7 mm chimney ($\bigtriangleup$), corresponding to a standard hole drilled through the wood, only E5 to G$\sharp$5 are in tune. 
This irregularity would make the instrument hard to play. 
Increasing the chimney to 20 mm ($\square$) improves tuning at the top of the register (G$\sharp$5 to C6), but worsens it elsewhere. 
A 13 mm chimney ($\bigcirc$) provides a better balance: E5 to A5 are in tune, and pitch errors remain acceptable higher up. 
This setup matches typical commercial clarinets. 
Still, B4 to C$\sharp$5 remain sharp, even in this compromise.

This tuning issue is well known to clarinetists \cite{nederveen1969acoustical}.
Manufacturers often address it by slightly flattening the corresponding notes in the first register, reducing the tuning gap across the register break  \cite{greenham2003clarinet}. 
Musicians are taught to compensate: they tighten the embouchure to raise the pitch of the lowest notes in the first register, and relax it to lower the pitch of the lowest notes in the second register. 
Some instruments even include extra keywork to correct the tuning of these problematic notes.

These tuning issue results from the constraints imposed by the \textit{throat} B$\flat$4. 
Beyond tuning, this note is not much appreciated among clarinetists, due to its inconsistent tone compared to the rest of the instrument \cite{mcginnis1943experimental, ICA_throattones_2016}. 
Musicians use correction fingerings, covering extra holes, to correct the pitch and tone of this note. 
Some clarinets even feature a special \textit{clear} B$\flat$4 mechanism that opens a hole dedicated to the B$\flat$4, instead of opening the RH \cite{hoeprich2008clarinet}.

Releasing the B$\flat$4 constraint significantly broadens the design space. 
Smaller hole diameters can still enable reliable register transitions, and they improve the tuning of the second register overall, consistent with Eq.~\eqref{chap5:eq:h}. 
Our optimal configuration is found for a 2.0 mm diameter, a 13 mm chimney, and a RH located 110 mm from the mouthpiece ($\times$). 
This setup cannot produce the \textit{throat} B$\flat$4 (the note is 69 cents flat), but it offers excellent tuning across the second register: the B4 is only 10 cents sharp. 
This solution assumes a dedicated B$\flat$4 hole is used instead.

Such a hole already exists on most clarinets, but its key is rarely used due to poor ergonomics. With our model now quantifying the full set of tuning constraints, it is up to instrument makers to make the \textit{clear} B$\flat$ truly playable.

\subsubsection{Relationships with linear approaches}

These results provide, for the first time, a quantitative prediction of the playability of the second register in the clarinet.
Previous attempts using traditional tools from nonlinear dynamics, such as continuation \cite{karkar2012oscillation,colinot2021multistability} and multistable regimes cartography \cite{maugeais2024makes,colinot2025cartography, tachibana2000sounding}, failed to identify RH configurations that fills the \textit{overblowing} criterion. This results from the absence of localized nonlinear losses, which play a key role in enabling register transitions.
By incorporating these losses, our model captures how geometry, control parameters, and regime selection influence each other.
It produces quantitative maps of suitable RH dimensions and positions.
In doing so, it complements and extends classical approaches, such as the expression of the length correction introduced by an open RH [Eq.~\eqref{chap5:eq:h}] \cite{benade1960mathematical, nederveen1969acoustical, hoekje1988abriefsummary, debut_analysis_2005}.

Results confirm the trends suggested by Equation~\eqref{chap5:eq:h}: tuning improves with smaller diameter and longer chimney; the lowest notes are the most out of tune (small value of $f_1^{(c)}$);  and tuning improves when the hole is located near the pressure node of the second modeshape, $L_1 = L/3$ (Fig.~\ref{chap5:fig:protocol:3:length} and \ref{chap5:fig:3}).
This approach has inherent limits. Equation~\eqref{chap5:eq:h} predicts perfect tuning for an infinitely small hole diameter or infinitely long chimneys. 
But in these extreme cases, the RH would no longer destabilize the first register, making it ineffective, as seen in Figure~\ref{chap5:fig:1} for $d_h=1.0$~mm, or in Figure~\ref{chap5:fig:1} for $d_h\leq1.5$~mm.

Further insight comes from the Bouasse-Benade prescription  \cite{bouasse1929tuyaux, benade1974nonlinear, gilbert2020minimal, campbell2021scientist}, which states that evenly spaced resonances enhance playability. 
Our simulations support this principle. 
Insufficient frequency shift of the first resonance (as in high notes, narrow and long-chimney holes) fails to destabilize the first register, as shown in Figure~\ref{fig:supp:lowLhF4} (Appendix.~\ref{chap5:fig:supp:sim}). 
Excessive detuning of both the first and second resonances (as in low notes, large diameter, short chimney and poorly-located hole) can lead to unstable behavior and favor either a higher register or the equilibrium, as shown in Figure~\ref{fig:supp:lowLhE3} (Appendix~\ref{chap5:fig:supp:sim}).

Our model captures all these behaviors and outlines the design region where transitions to the second register are possible.
The quantification of these boundaries fills the gaps left by earlier models.
Nonlinear losses are essential to this predictive ability.

\subsection{Concluding remarks}
We introduced a sparse clarinet model that reproduces the observed register transitions caused by the opening of the register hole. From this model, we define a range of register hole dimensions that ensure second register production across the entire playing range.
This design space becomes markedly narrower if the register hole must also raise the A4 by a semitone to produce the \textit{throat} B$\flat$4. 
Under this constraint, the only way to mitigate the critical tuning issues of second register is to lengthen the register hole by inserting a metal tube into the wood. 
Relaxing this requirement enables to improve the tuning of the second register in modern instruments.

This is, to our knowledge, the first physical model that quantitatively predicts the playability limits of the second register, enabled by the inclusion of localized nonlinear losses. Previous approaches relying on the linear response of the resonator were not sufficient to identify suitable register hole configurations.

The implications of the model extend beyond clarinets. Localized nonlinear losses occur in all reed instruments with register holes. This modeling approach can thus be applied to oboes, saxophones, bassoons and their relatives.

The results provide a quantitative foundation for clarinet design, highlighting clear trade-offs between tuning, playability, and chromatic continuity. They also suggest a practical path forward: using a separate B$\flat$4 hole and improving its ergonomics could unlock better tuning of the second register without sacrificing chromaticism.
By accounting for nonlinear losses, we offer not just a better model, but a step toward more precisely tuned and more playable woodwind instruments.

\subsection*{Acknowledgments}
The authors warmly thank P.\ Bindzi and V.\ Long for producing the clarinet prototypes; L.\ Maignan, G.\ Gatti, A.\ Patsinakidou, P.\ F.\ Oliveira and H.\ Arzumanyan for carrying out the playing tests. 
This study has been supported by the French ANR LabCom LIAMFI (ANR-16-LCV2-007-01). 

\subsection*{Materials and Correspondence}
The data and methods that support this study are available in \url{https://doi.org/10.5281/zenodo.16169580}.

\vfill ~

\appendix
\makeatletter
\@addtoreset{figure}{section}              
\makeatother
\renewcommand{\thefigure}{\thesection.\arabic{figure}}

\section{Supplementary data: Input Impedance measurements}\label{chap5:sec:Zin}
\begin{figure}[H]
	\centering
	\includegraphics[width=\textwidth]{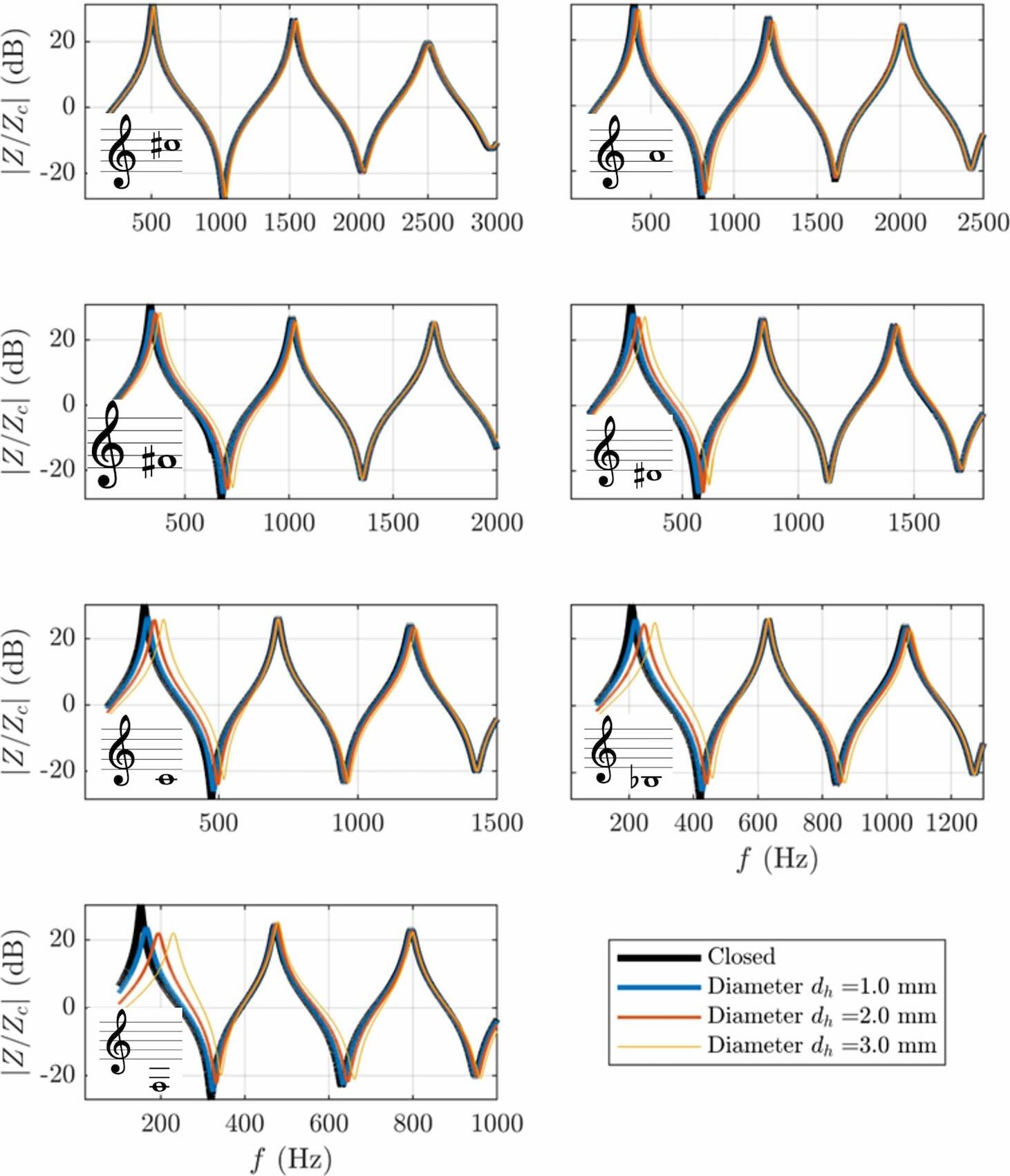}
	\caption{Amplitude of the measured normalized input impedances. Each panel refers to one same note (i.e.\ the same lower body). }
	\label{fig:impedance}
\end{figure}

\section{Supplementary data: Playing tests statistics}\label{chap5:sec:stats}
\begin{figure}[H]
	\centering
	\includegraphics[width=.8\textwidth]{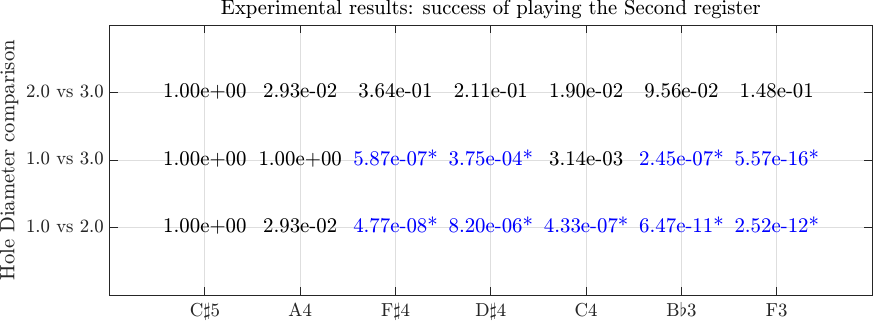}
      \caption{\textbf{Statistical comparison of the success of playing the second register.} 
	Pairwise $p$-values from one-sided Fisher's exact tests between two diameter sizes, with the alternative hypothesis $\mathrm H_1$: ``The second register is more produced for diameter $j$ (second value on the y-label) than diameter $i$ (first value on the y-label)''.
	Values in blue color indicate rejection of $H_0$.
	The significance level is Bonferroni corrected to $p<0.05/(7 \cdot 3 \cdot 4)=7.94\cdot 10^{-4}$.
	The total number of trials per note and diameter is $N=78$.}
	\label{fig:stats1}
\end{figure}

\begin{figure}[H]
	\centering
	\includegraphics[width=.9\textwidth]{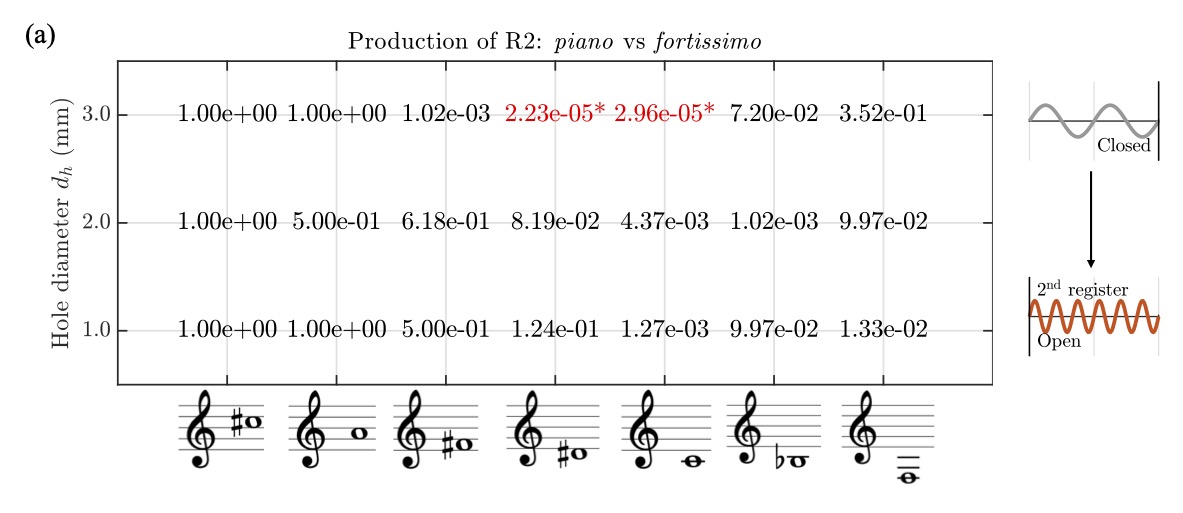}
	\includegraphics[width=.9\textwidth]{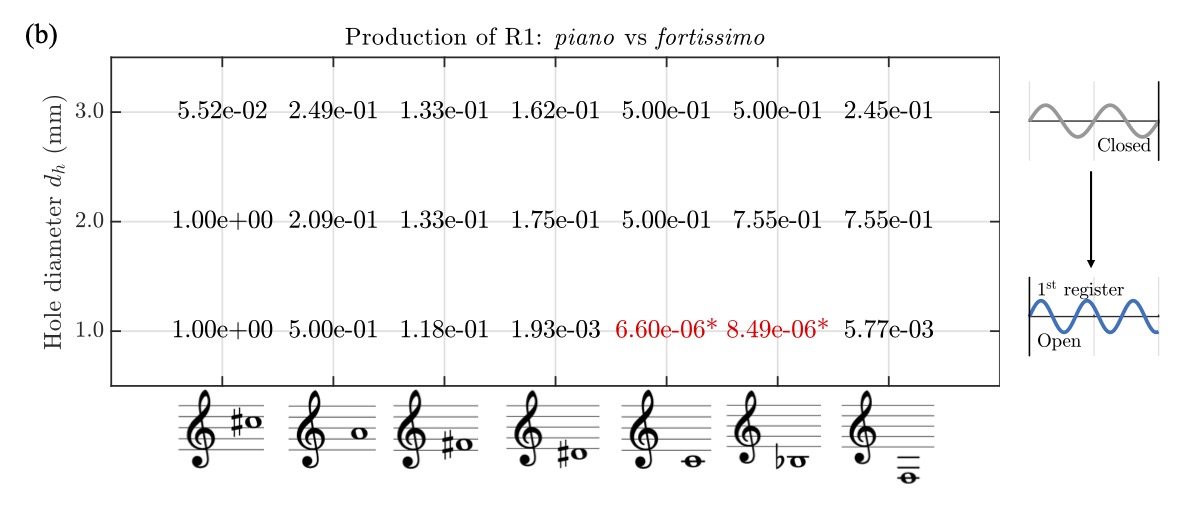}
	\caption{\textbf{Statistical comparison of the production success of a given regime.}
	 Pairwise $p$-values from one-sided Fisher's exact tests comparing \textit{piano} vs.~ \textit{fortissimo} nuances, with the alternative hypothesis $\mathrm H_1$: ``the register of interest is more successfully produced in \textit{fortissimo} than in \textit{piano}''.
	Red values indicate significance.
	The significance level is Bonferroni corrected to $p<0.05/(7 \cdot 3 \cdot 3)=7.94\cdot 10^{-4}$.
	The total number of trials per nuance is $N=26$.
    Subfigures (a) and (b) show results for the second (R2) and first (R1) registers, respectively.}
	\label{fig:stats2}
\end{figure}

\section{Supplementary data: Simulation extended results}\label{chap5:fig:supp:sim}
\begin{figure}[H]
	\centering
	\includegraphics[width=\textwidth]{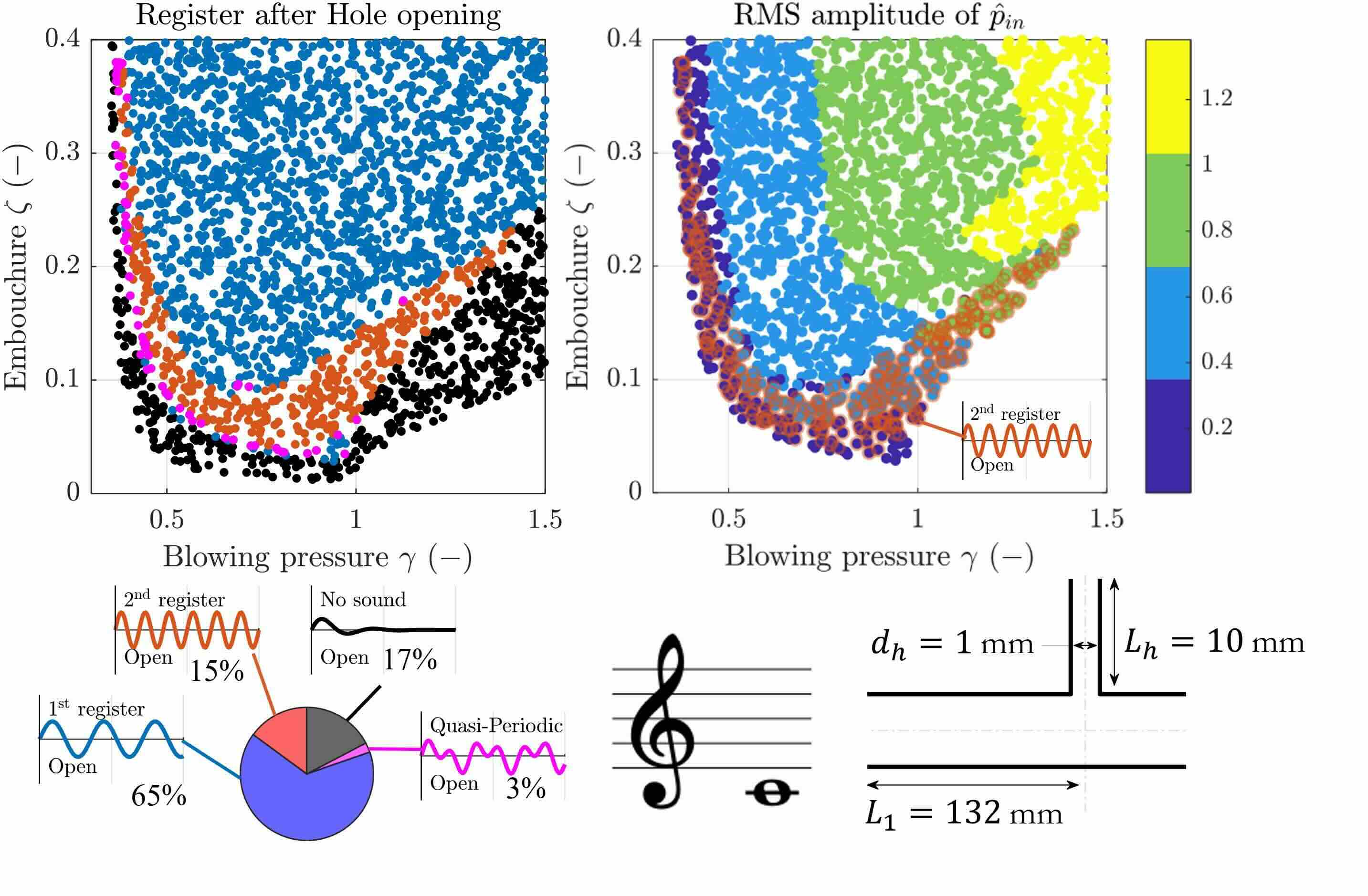}
	\caption{
	\textbf{The second register is mostly played at soft nuances for small-diameter register holes.} Simulation result: Influence of the amplitude of the input acoustic pressure on the register played for a thin-diameter register hole.
	Hole openings are simulated for note C4 with $d_h=1$~mm.
	Results show a high proportion of the first register (65\%), and a weak proportion of the second register (15\%).
	The right figure shows that the second register is mostly produced for low RMS amplitudes of the input acoustic pressure.
	Thus, the second register is mostly played for soft nuances.}
	\label{fig:supp:sim:nuance}
\end{figure}

\begin{figure}[H]
	\centering
	\includegraphics[width=\textwidth]{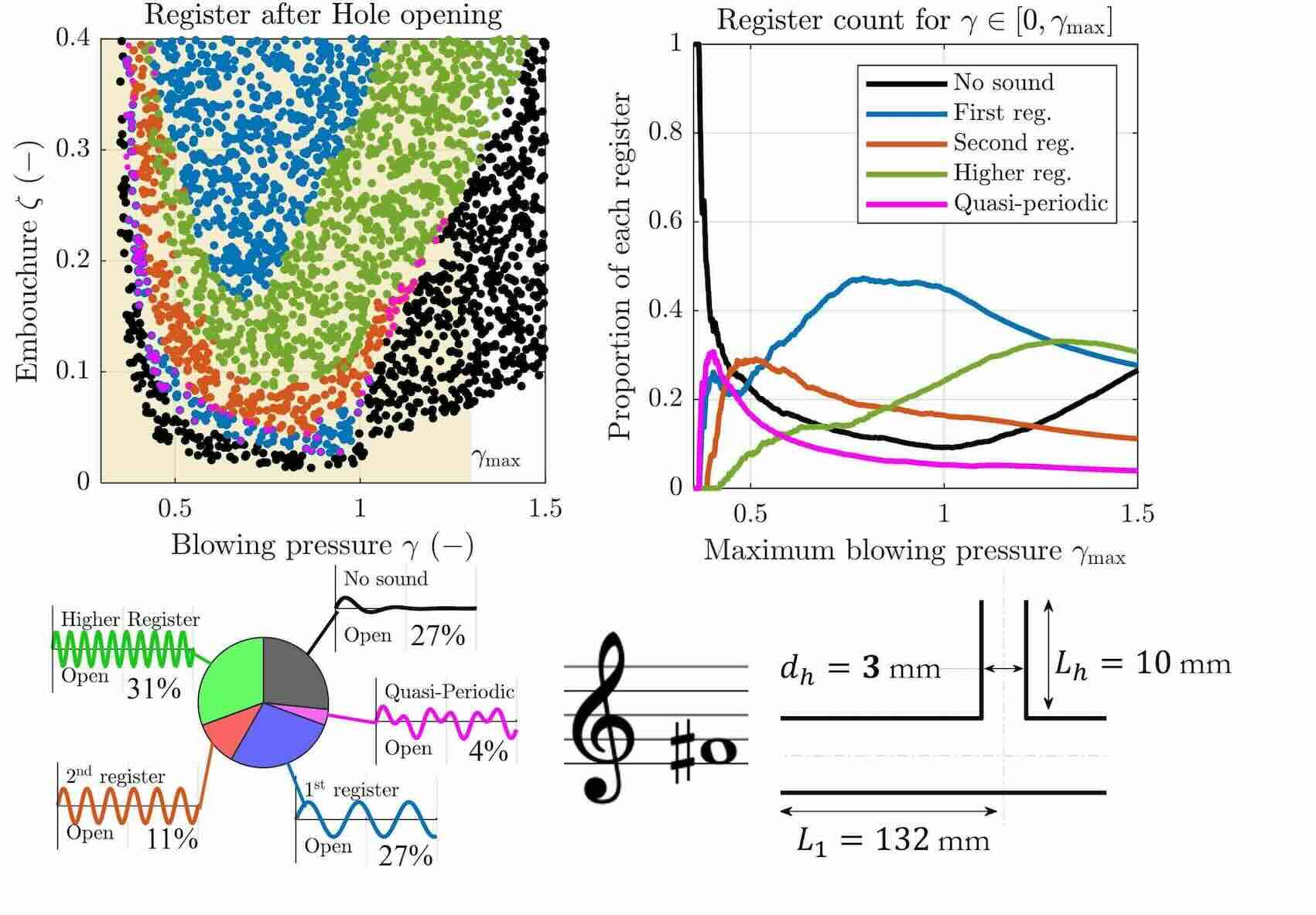}
	\caption{
	\textbf{Higher registers are favored for very high blowing pressures.} Evolution of the proportion of  the different registers with respect to the size of the blowing pressure window $[0, \gamma_\mathrm{max}]$.
	Simulation for note F$\sharp4$ with a register hole of diameter $d_h=3.0$~mm.
	The first register is produced in majority for $0.54<\gamma_\mathrm{max}<1.26$.
	These results underline the strong representation of higher registers at high blowing pressure.
	}
	 \label{fig:supp:gmax}
\end{figure}

\begin{figure}[H]
	\centering
	\includegraphics[width=.8\textwidth]{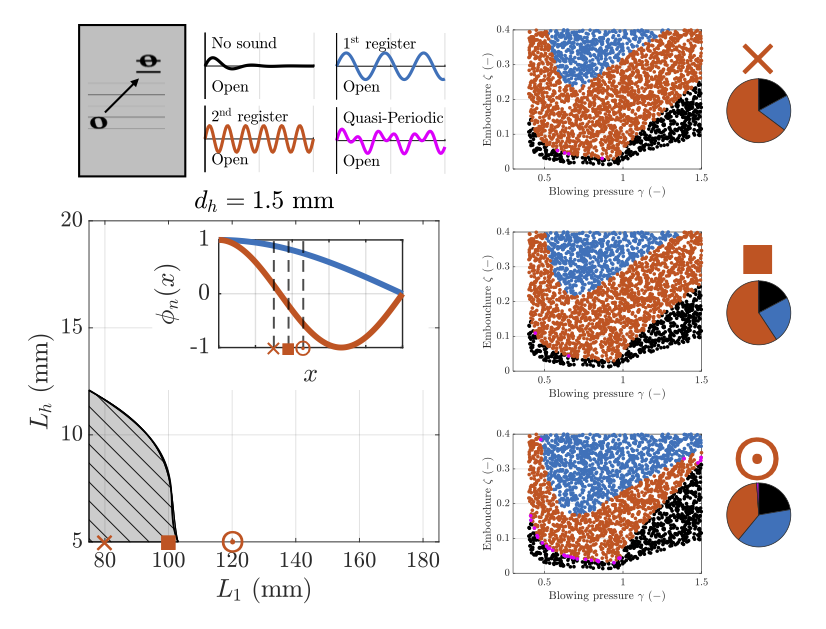}
	\caption{
	\textbf{Weak frequency shift on the first mode reduces second-register emergence.}
	Note F4 is considered ($f_1=320$~Hz).
	The register hole has a very low diameter $d_h=1.5~$mm, and a short chimney height $L_h=5$~mm.
	Three hole relative positions are considered: $x_h=\{0.30, 0.38, 0.46\}$. 
	The amplitude of the first mode shape $\phi_1(x_h)$ decreases monotonically as $x_h$ increases. 
	As a result, following the inharmonicity Equation \eqref{chap5:eq:h}, the frequency shift on the first resonance decreases as $x_h$ increases.
	Simulation highlight an increase of stable first register in the control parameters space, making the second register harder to produce.
	}
	\label{fig:supp:lowLhF4}
\end{figure}
\begin{figure}[H]
	\centering
	\includegraphics[width=.8\textwidth]{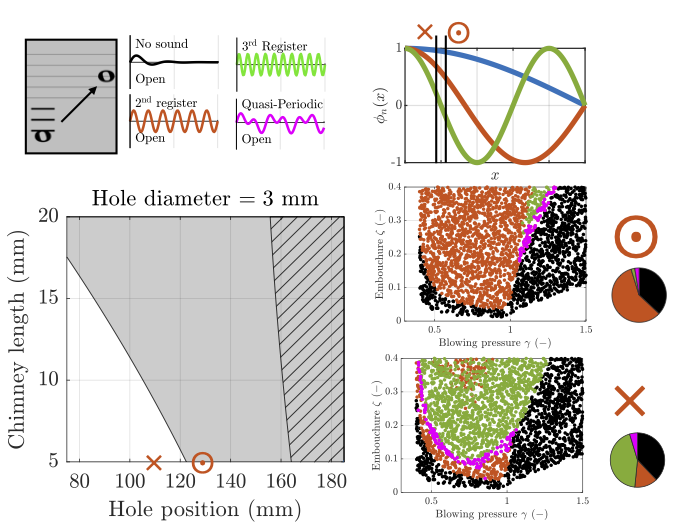}
	\caption{
	\textbf{Extreme frequency shift of the first and second resonance frequencies.}
	We consider note E4 ($f_1=149$~Hz).
	The register hole has a diameter $d_h=3.0~$mm, and a short chimney height $L_h=5$~mm.
	Two hole relative positions are considered: $x_h=\{0.197, 0.224\}$. 
	For these positions, $\phi_1$ and $\phi_2$ have a high amplitude, whereas $\phi_3(x_h)$ is close to zero.
	We also note that the amplitude of $\phi_2$ increases as the hole becomes closer to the mouthpiece.
	As a result, following the inharmonicity Equation \eqref{chap5:eq:h}, the first two resonance frequencies are heavily shifted.
	Simulations highlight a dramatic decrease of stable second register as the hole becomes closer to the mouthpiece.
	}	
	\label{fig:supp:lowLhE3}
\end{figure}

\bibliography{biblio.bib}

\end{document}